\documentclass[fleqn,usenatbib,useAMS]{mnras}

\usepackage{newtxtext,newtxmath}
\usepackage[T1]{fontenc}
\usepackage{orcidlink}

\DeclareRobustCommand{\VAN}[3]{#2}
\let\VANthebibliography\thebibliography
\def\thebibliography{\DeclareRobustCommand{\VAN}[3]{##3}\VANthebibliography}

\usepackage{graphicx}	% Including figure files
\usepackage{amsmath}	% Advanced maths commands
\title[SN~2015da]{SN~2015da:  Late-time observations of a persistent superluminous Type~IIn supernova with post-shock dust formation}

\author[Smith et al.]{Nathan Smith \orcidlink{0000-0001-5510-2424},$^1$\thanks{Email:
    nathans@as.arizona.edu}  Jennifer E. Andrews \orcidlink{0000-0003-0123-0062},$^2$  Peter Milne,$^1$  Alexei V. Filippenko \orcidlink{0000-0003-3460-0103},$^{3}$ Thomas G. Brink\orcidlink{0000-0001-5955-2502},$^{3,4}$ \newauthor Patrick L.\ Kelly\orcidlink{0000-0003-3142-997X},$^{5}$ Heechan Yuk\orcidlink{0000-0002-7720-3418},$^{6}$ and Jacob E.\ Jencson\orcidlink{0000-0001-5754-4007}$^{7,8}$ \\  %Weikang Zheng,$^{3,5}$ 
     $^1$Steward Observatory, University of Arizona, 933 North Cherry Avenue, Tucson, AZ 85721, USA \\ 
     $^2$ Gemini Observatory, 670 N. Aohoku Place, Hilo, Hawaii, 96720, USA \\  
    $^3$ Department of Astronomy, University of California, Berkeley, CA 94720-3411, USA \\
    $^4$ Wood Specialist in Astronomy \\
    $^5$ School of Physics and Astronomy, University of Minnesota, 116 Church Street SE, Minneapolis, MN 55455, USA \\ %Eustace Specialist in Astronomy 
    $^6$ Homer L. Dodge Department of Physics and Astronomy, University of Oklahoma, Norman, OK 73019, USA \\
    $^7$ Department of Physics and Astronomy, Johns Hopkins University, 3400 North Charles Street, Baltimore, MD 21218, USA \\ 
    $^8$ Space Telescope Science Institute, 3700 San Martin Drive, Baltimore, MD 21218, USA
    }
  
\date{Accepted XXX. Received YYY; in original form ZZZ}

\pubyear{2023}

\begin{document}
\label{firstpage}
\pagerange{\pageref{firstpage}--\pageref{lastpage}}
\maketitle

\begin{abstract}
We present photometry and spectroscopy of the slowly evolving
superluminous Type~IIn supernova (SN) 2015da.  SN~2015da is
extraordinary for its very high peak luminosity, and also for
sustaining a high luminosity for several years. Even at 8\,yr after explosion, 
SN~2015da remains as luminous as the peak of a normal SN~II-P.  The total 
radiated energy integrated over this time period (with no bolometric
correction) is at least $1.6 \times 10^{51}$ erg (or 1.6\,FOE).  
Including a mild bolometric correction, adding 
kinetic energy of the expanding cold dense shell of swept-up circumstellar 
material (CSM), and accounting for asymmetry, the total explosion kinetic 
energy was likely 5--10\,FOE.  Powering the light curve with CSM
interaction requires an energetic explosion and 20\,M$_{\odot}$ of H-rich CSM, which
in turn implies a massive progenitor system $>$30\,M$_{\odot}$.
Narrow P~Cyg features show
steady CSM expansion at 90\,km\,s$^{-1}$, requiring a high average
mass-loss rate of $\sim$0.1\,M$_{\odot}$\,yr$^{-1}$ sustained for 2 centuries before
explosion (although ramping up toward explosion time).  No
current theoretical model for single-star pre-SN mass loss can account
for this. The slow CSM, combined with broad wings of H$\alpha$ indicating H-rich material in the unshocked ejecta, disfavour a pulsational pair instability model for the pre-SN mass loss.  Instead, violent pre-SN binary interaction is a likely cuprit.
Finally, SN~2015da exhibits the characteristic asymmetric
blueshift in its emission lines from shortly after peak until
the present epoch, adding another well-studied superluminous SNe~IIn
with unambiguous evidence of post-shock dust formation.
\end{abstract}


\section{INTRODUCTION}

Since the first well-observed example of SN~2006gy, superluminous
supernovae (SLSNe) have presented a significant challenge for stellar
evolution theory.  Hydrogen-deficient examples \citep{quimby11},
usually referred to as SLSNe~Ic, show broad lines in their spectra and
have been modeled as the result of post-explosion energy deposition by
a magnetar wind \citep{woosley10,kb10}.  On the other hand, SLSNe with
hydrogen lines usually exhibit relatively narrow components in their line profiles,
and are thus classified as SLSNe~IIn. These, like
their more modest-luminosity analogs (i.e., regular SNe~IIn; \citealt{schlegel90,filippenko97}), 
are thought to be powered by shock interaction
with dense circumstellar material (CSM).  Key challenges for
stellar evolution posed by SLSNe~IIn are getting massive
progenitors to evolve up to the time of core collapse with their H
envelope intact, then shedding much of that H envelope in
sudden bursts of mass loss just prior to core collapse, and finally
getting massive progenitors to explode successfully with higher than
average energy \citep[see][]{smith07,smith10,groh13}.

CSM interaction can produce wide diversity in the observed properties
of SNe~IIn, since any type of SN explosion can, in principle, have
dense CSM, and that CSM can vary in its mass, radial distribution, and
geometry \citep[see][for a review of interacting SNe]{smith17}.  Some
have only a small additional luminosity, whereas others become
extraordinarily luminous (as in the case of SLSNe~IIn).  Some have
CSM interaction signatures that are fleeting, fading in only a few
days (like the recent example of SN~2023ixf in M101;
\citealt{wynn23,smith23,azalee23}), and others have extremely strong
interaction that can persist for decades (as in the prolonged cases of
SN~1988Z and SN~2005ip;
\citealt{smith17sn05ip,smith09sn05ip,fox10,williams02}).  Large fluctuations 
in pre-SN mass loss may lead to highly variable CSM interaction, observed as
bumps or dips in the late-time light curves \citep{nyholm17,smith17sn05ip}.
CSM interaction can also lead to efficient dust formation in the rapidly
cooling post-shock layers, indicated by a combination of excess infrared (IR)
emission and blueshifted emission-line profiles 
\citep{smith08jc,smith09sn05ip,smith12,gall14,smith20}.  Ongoing
CSM interaction can cause SNe~IIn to appear as luminous IR
sources for many years, as pre-shock CSM dust is continually heated by
the advancing shock \citep{fox13,fox13gl,fox15,fox20}.  Spectral
signatures of dust formation are discussed later in this paper.

SNe~IIn require astounding progenitor mass loss to produce their dense
CSM.  The least of these require mass loss at around $10^{-3}$\,M$_{\odot}$\,yr$^{-1}$, 
which would overlap with the strongest known examples of
steady stellar winds, like the current wind of $\eta$~Carinae
\citep{smith03,hillier06} or those of the most extreme known red
supergiants (RSGs) like VY~CMa \citep{smith09,smith09sn05ip,decin06}.  Normal
RSG winds are 100--1000 times weaker \citep{beasor20}.  The majority
of SNe~IIn require even higher mass-loss rates of $10^{-2}$
to $10^{-1}$\,M$_{\odot}$\,yr$^{-1}$ or more, far beyond the limiting
capability of any known radiatively driven steady stellar wind
\citep{so06,smith14}.  This points instead to episodic, eruptive, and
explosive mass-loss mechanisms.  The only well-established observed
precedent for this mode of mass loss is the giant eruptions of
luminous blue variables (LBVs; \citealt{smith11lbv}).  The mechanism for
these eruptions remains debated, but may arise from super-Eddington
continuum-driven winds \citep{so06,owocki04,owocki17,quataert16} or violent binary
interaction \citep{smith18,st06}.  There are many cases of SNe~IIn and
SLSNe~IIn in between the observed extremes, representing a continuum
in CSM properties (e.g., \citealt{dickinson23}).

In SLSNe~IIn, the luminosity peaks at early times are likely
powered by diffusion of shock-deposited energy in an opaque CSM
envelope, analogous to a delayed shock breakout with photons degraded
to optical wavelengths.  This was first proposed to explain the main
peak of SN~2006gy \citep{sm07}, and was subsequently adopted more
generally for SNe~IIn \citep{bl11,ci11}. This mechanism leads to
extremely high efficiency for converting kinetic energy into radiation
\citep{sm07,vm10}.  At later times when the optical depth drops, we
can see the shock interaction more directly, and SNe~IIn can be
luminous in X-rays that escape the shock-interaction zone
\citep{chandra09,chandra12,chandra22,pooley02,tsuna21}.  

In order to account for the high
luminosity of SLSNe through CSM interaction, the required total amount
of CSM mass in SLSNe is extreme, typically requiring 5--10\,M$_{\odot}$
(sometimes even 20--25\,M$_{\odot}$ or more) of CSM ejected by the progenitor
shortly before exploding
\citep{dickinson23,nicholl20,rest11,sm07,smith08tf,smith10,woosley07}.
When one factors in the mass of the compact remnant, the SN ejecta,
and mass lost by winds during the star's lifetime, this mass budget
points to quite massive progenitor stars for some SLSNe~IIn, in some
cases with intial masses that must exceed 40--50\,M$_{\odot}$.
High-mass progenitors retaining their H envelopes until shortly before
death, when they are ejected in LBV-like eruptions, directly
contradicts predictions of traditional stellar-evolution models.  Such
stars are expected to lose their H envelopes in winds during their
lifetime, perhaps passing through a transitory LBV phase, before
becoming H-free Wolf-Rayet (WR) stars that die as SNe~Ibc
\citep{cm86,woosley93,mm00,heger03}.  Some of this disagreement may be
alleviated by reduced  wind mass-loss
rates \citep[see the review by][for an extended
  discussion]{smith14}.  These more realistic lower mass-loss rates
have, however, not yet percolated into the most commonly used stellar
evolution model grids.  Even if more realistic lower mass-loss rates
allow massive stars with $M_{\rm ZAMS} \ga 30\,{\rm M}_{\odot}$ to retain
their H envelopes until death, their successful energetic explosion
and their bursts of pre-SN mass loss still pose a theoretical challenge.

The question of how long the strong CSM interaction lasts is
particularly important for diagnosing the underlying physical mechanism(s) that
may have produced dense CSM.  An observed estimate for
the expansion speed of the CSM can help constrain the time
period before core collapse when the CSM was ejected.  Some proposed 
mechanisms for episodic pre-SN mass loss include
energy transferred to the envelope by wave driving in advanced nuclear
burning phases \citep{qs12,sq14,fuller17,fr18,wf21}, the pulsational pair
instability or other late-phase burning instabilities
\citep{woosley07,woosley17,am11,sa14,renzo20}, or an inflation of the
progenitor's radius (perhaps caused by the previous mechanisms) that
triggers violent binary interaction like collisions or mergers before
core collapse \citep{sa14} or mergers with compact companions
\citep{fw98,schroder20}.  Of these various mechanisms, only the ones with binary interaction predict highly
asymmetric distributions of CSM (disk-like or bipolar), relevant to
asymmetric line-profile shapes and high polarization seen in SNe~IIn
\citep{bilinski23}.  In terms of timescale, wave driving only operates
for about 1\,yr during Ne and O burning, and is therefore too short to
account for SNe~IIn and SLSNe~IIn, which require heavy mass loss for
decades or centuries.  The pulsational
pair instability also operates primarily during O burning, although
its timescale can be extended significantly in some cases owing to
Kelvin-Helmholtz relaxation after a powerful pulse
\citep{woosley17,renzo20}. However, the pair instability should only
operate in extremely massive stars and is therefore too rare to
account for SNe~IIn that make up $\sim$8 \% of ccSNe
\citep{smith11frac}.  It is a good candidate to explain the
pre-SN mass loss for some of the rare SLSNe~IIn \citep{woosley07}, 
although as we argue below, not for SN~2015da.

\begin{figure}
\begin{center}\includegraphics[width=2.5in]{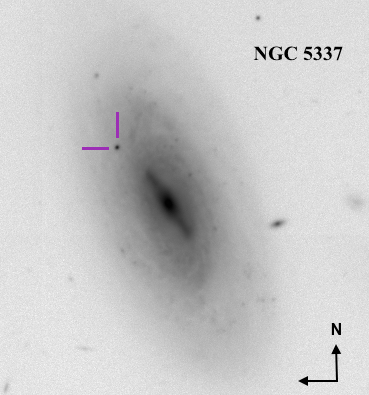}
\end{center}
\caption{A late-time LBT/MODS $r$-band image of SN~2015da and its barred spiral host galaxy NGC~5337 taken on 2018 January 16 (day 1104).  The SN position is indicated by the violet crosshairs.}
\label{fig:finder}
\end{figure}

\begin{table*}\begin{center}\begin{minipage}{4.25in}
      \caption{SLOTIS $BVRI$ photometry}
\centering
\small
\begin{tabular}{@{}ccccccccccc}\hline\hline
MJD & $B$ & $\sigma$ & $V$  & $\sigma$ & $R$ & $\sigma$ & $I$ &$\sigma$ \\
  &(mag) &(mag) &(mag) &(mag) &(mag) &(mag) &(mag) &(mag) \\
\hline
%% Begin data
57148.7 &17.87 &0.23  &16.56 &0.06  &15.66 &0.03  &15.02  &0.04 \\
57150.7 &17.60 &0.21  &16.57 &0.07  &15.69 &0.05  &15.05  &0.04 \\
57152.8 &17.66 &0.15  &16.56 &0.08  &15.72 &0.04  &15.03  &0.03 \\
57162.7 &17.34 &0.30  &16.71 &0.10  &15.79 &0.05  &15.12  &0.06 \\
57168.7 &17.80 &0.14  &16.84 &0.09  &15.89 &0.02  &15.16  &0.04 \\
57171.7 &17.88 &0.27  &16.73 &0.12  &15.87 &0.03  &15.15  &0.05 \\
57174.7 &17.24 &0.37  &16.80 &0.14  &15.91 &0.03  &15.21  &0.05 \\
57177.7 &17.67 &0.21  &16.96 &0.11  &15.94 &0.03  &15.18  &0.05 \\
57192.7 &...   &...   &16.92 &0.14  &15.97 &0.04  &...    &... \\
57195.7 &...   &...   &16.85 &0.11  &16.02 &0.05  &...    &... \\
57387.0 &...   &...   &...   &...   &16.84 &0.04  &16.43  &0.09 \\
57441.9 &19.13 &0.20  &...   &...   &17.15 &0.06  &16.89  &0.04 \\
57825.9	&...   &...   &...   &...   &18.39 &0.08  &...    &... \\
57902.7	&...   &...   &...   &...   &18.55 &0.06  &...    &... \\
%% End data
\hline \\
\end{tabular}
\label{tab:slotis}
\end{minipage}\end{center}
\end{table*}

%%%%%%%

In this paper, we discuss the long-lasting superluminous Type IIn event SN~2005da (also known as PSN J13522411+3941286), discovered in the barred spiral NGC~5337 (see Fig.~\ref{fig:finder}).  The photometric and spectroscopic evolution have already been discussed in detail by \citet[][T20 hereafter]{tar20}, and we refer the reader to that paper for background information about its discovery, host galaxy and environment, and early-time evolution.  To be consistent with T20, we adopt the same values for the distance $d=53.2$\,Mpc, host-galaxy redshift $z=0.0071$, distance modulus $m-M=33.63$\,mag, line-of-sight reddening $E(B-V)=0.98$\,mag (combined 0.01\,mag Galactic and 0.97\,mag host-galaxy reddening), and metallicity $\sim 0.6$\,Z$_{\odot}$.  For the sake of comparison, we also adopt the T20 explosion UTC date of 2015 Jan. 8.45 (JD = 2,457,030.95), which was well constrained by observations to be about 1.5\,days before the first detection.  This makes SN~2015da an unusual case of being a nearby superluminous SN~IIn discovered within a few days of explosion while still in its early rise, similar to SN~2006gy \citep{smith07}, but closer to us.  Another well-studied SN~IIn, SN~2010jl, was even closer at $\sim$49\,Mpc \citep{smith11} and spectroscopically similar to SN~2015da (T20), but discovered later in its evolution around the time of peak brightness \citep{stoll11}.  Compared to T20, we present an independent and complementary set of photometric and spectroscopic data that shows similar overall evolution, so our discussion of the observations and results in Section 2 and 3 are brief.  However, a difference is that we include a series of higher-resolution spectra over a longer time, and we also include photometry extended to later times.  In our analysis, we briefly discuss areas where our interpretation is complementary to that of T20 as we discuss the overall energy and mass budget of SN~2015da, but we also highlight areas where our analysis and interpretation differ from those of T20, particularly in the interpretation of the emission-line-profile evolution.

\section{OBSERVATIONS}

\subsection{Imaging} \label{obs:slotis}
After discovery of SN~2015da, we added its field to the queue of the robotic Super-LOTIS 24\,inch telescope \citep[SLOTIS;][]{Wil08} on Kitt Peak for multifilter ($B$, $V$, $R$, and $I$) follow-up observations. The seeing varied in the range $\sim$2$-$4 arcsec. Images were automatically reduced and calibrated using a custom pipeline, % by P.\ Milne, 
and aperture photometry was performed manually. 
%The magnitudes were calibrated using the reference star list from the SLOTIS pipeline, shown in Table 4. 
Table \ref{tab:slotis} lists the resulting SLOTIS photometry.  We also obtained a series of $B$, $V$, and $R$ images using the Mont4k CCD on the 61\,inch Kuiper telescope on Mt. Bigelow \citep{Fon14}; the seeing was typically 1$-$2 arcsec.  Aperture photometry was performed manually. Some of the reference stars were outside the field of view or saturated, and were thus excluded from the reduction. Table \ref{tab:kuiper} summarizes the results.  

\begin{table}\begin{center}\begin{minipage}{3.25in}
      \caption{Kuiper/Mont4k $BVR$ photometry}
\centering
\small
\begin{tabular}{@{}lcccccc}\hline\hline
MJD & $B$ & $\sigma$ & $V$  & $\sigma$  & $R$ & $\sigma$ \\ 
 &(mag) &(mag) &(mag) &(mag)  &(mag) &(mag) \\
\hline
%% Begin data
57166.8 &...   &...   &16.79	&0.07 &15.88 &0.06\\
57465.9 &19.16 &0.08  &18.29	&0.08 &17.19 &0.04\\
57495.9 &19.26 &0.08  &18.41	&0.08 &17.38 &0.03\\
57521.8 &19.31 &0.08  &18.56	&0.06 &17.52 &0.02\\
57543.7 &19.54 &0.07  &18.62	&0.10 &17.64 &0.04\\
57578.7 &...   &...   &18.72	&0.08 &17.77 &0.02\\
57864.9 &...   &...   &...    &...  &18.48 &0.05 \\
57923.8 &...   &...   &...    &...  &18.59 &0.04 \\
58109.0 &...   &...   &...    &...  &18.73 &0.05 \\
%% End data
\hline \\
\end{tabular}
\label{tab:kuiper}
\end{minipage}\end{center}
\end{table}

\begin{table}\begin{center}\begin{minipage}{3.25in}
      \caption{Late-time optical photometry}
\centering
\begin{tabular}{@{}lcccccc}\hline\hline
UTC Date &Age (days) &Tel./Instr. &filt. &mag   &$\sigma$ \\
\hline
%% Begin data
2018-01-16  &1104 &LBT/MODS & $g$ &20.17 &0.07  \\
2018-01-16  &1104 &LBT/MODS & $r$ &19.28 &0.03  \\
2019-06-01  &1605 &MMT/Bino & $r$ & 19.47 & 0.05  \\
2022-03-03  &2582 &LBT/MODS & $g$ & 21.19 & 0.10 \\
2022-03-03  &2582 &LBT/MODS & $r$ & 20.21 &0.10  \\
2022-03-25  &2604 &MMT/Bino & $g$ & 21.09 & 0.09  \\
2022-03-25  &2604 &MMT/Bino & $r$ & 20.15  & 0.07   \\
2022-03-25  &2604 &MMT/Bino & $i$ & 20.38 &  0.10 \\
2022-05-06  &2646 &LBT/MODS & $g$ & 21.17 & 0.12 \\
2022-05-06  &2646 &LBT/MODS & $r$ & 20.18 & 0.08 \\
%% End data
\hline \\
\end{tabular}
\label{tab:latephot}
\end{minipage}\end{center}
\end{table}

\begin{table}\begin{center}\begin{minipage}{3.25in}
      \caption{Near-infrared photometry}
\centering
\scriptsize
\begin{tabular}{@{}ccccccccc}\hline\hline
MJD &Age    &Tel. & $J$ & $\sigma_J$ & $H$ & $\sigma_H$ & $K$ & $\sigma_K$ \\
    &(days) & &(mag) &(mag) &(mag) &(mag) &(mag) &(mag) \\
\hline
%% Begin data
57144 &114 &UKIRT  &14.07 &0.01 &13.63 &0.01  &13.27 &0.02 \\
57178 &148 &UKIRT  &14.41 &0.01 &13.94 &0.01 &13.46 &0.02 \\
57193 &163 &UKIRT  &14.52 &0.01 &14.06 &0.01 &13.55 &0.02 \\
%58242 &274 &MMT & & & & & & \\
%58655 &687 &MMT & & & & & & \\
%% End data
\hline \\
\end{tabular}
\label{tab:ukirt}
\end{minipage}\end{center}
\end{table}

%MJD      Filter  ApMag  ApMag_e  PSFMag  PSFMag_e
%58242.2  J       18.94  0.10     18.09   0.08
%58242.2  H       17.32  0.11     17.21   0.11
%8655.3  J       20.12  0.13     20.02   0.07
%58655.3  H       18.75  0.11     18.59   0.09
%58655.3  K       16.44  0.15     16.37   0.15

Late time $g$-, $r$-, and $i$-band images were also obtained between 2018 to 2022 using the imaging mode of the Multi Object Double Spectrograph \citep[MODS]{Bya00} mounted  on the 2$\times$8.4\,m Large Binocular Telescope (LBT), as well as the imaging mode of Binospec  \citep{fabricant19} on the 6.5\,m MMT.  An LBT/MODS $r$-band image is shown in Figure~\ref{fig:finder}.   
%From these LBT/MODS images, we measure $g$=20.17 ($\pm$0.07) mag and $r$=19.28 ($\pm$0.03) mag.  
Photometry measured from these late-time MMT and LBT images is listed in Table~\ref{tab:latephot}.

\begin{figure*}
\includegraphics[width=7.0in]{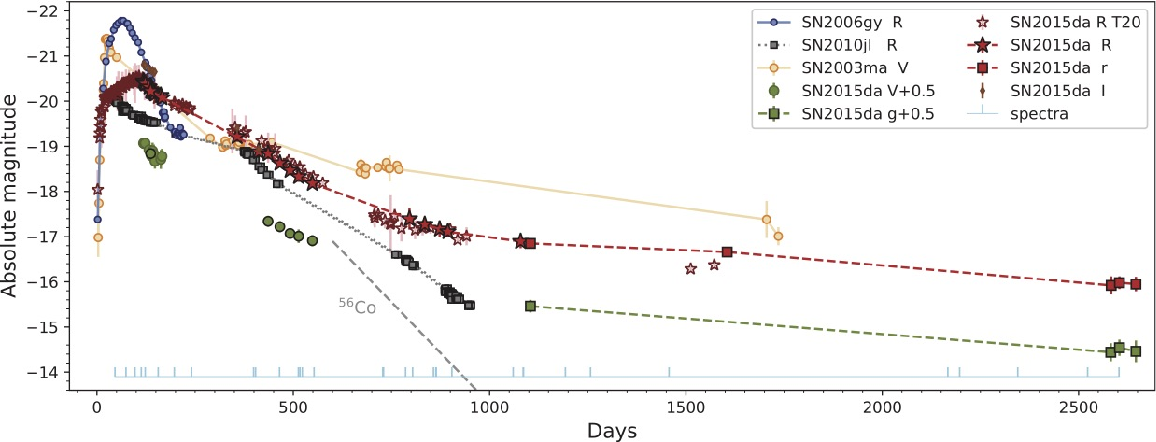}
\caption{Absolute magnitude light curve of SN~2015da compared to other SLSNe~IIn.  $R$-band light curve of SN~2006gy \citep{smith07} in blue circles and SN~2010jl \citep{fransson14} in grey/black squares, plus the $V$-band light curve of SN~2003ma in orange \citep{rest11}, are given for comparison.  The $R$ light curve of SN~2015da from T20 is shown with light-red unfilled stars and a dashed line.  Our new SN~2015da photometry is plotted for $V$ (green solid circles), $R$ (red stars), and $I$ (brown diamonds).  Photometry from Kuiper images has black outlines, while SLOTIS points do not. For the Kuiper and SLOTIS $R$ photometry, we have subtracted a constant baseline flux to correct for galaxy light in the aperture.   The late-time LBT/MODS and MMT/Binospec photometry in $g$ and $r$ is shown with squares, but using the same fill colours as $V$ and $R$.  The $^{56}$Co decay rate with an arbitrary luminosity is also shown for comparison (grey dashed line), and times of our spectra are indicated with light-blue tick marks at the bottom.}
\label{fig:lightcurve}
\end{figure*}

Figure~\ref{fig:lightcurve} shows our optical $V/g$, $R/r$, and $I$-band photometry from SLOTIS, KUIPER, and LBT converted to absolute magnitudes using the adopted distance and reddening corrections noted in the Introduction.  For comparison, we also show the $R$-band photometry of SN~2015da from T20, as well as SLSN~IIn $R$ light curves of SN~2006gy \citep{smith07} and SN~2010jl \citep{fransson14}.

Three sets of near-IR $JHK$-band images were obtained during the same period as the SLOTIS data, using the UK Infrared Telescope (UKIRT) Wide Field Camera instrument \citep[WFCAM;][]{Hod09}. The seeing, estimated from the full width at half-maximum intensity (FWHM) of stars on the CCD frame, varied from $1''$ to $2''$. Aperture photometry was performed manually, and the magnitudes were calibrated using the same reference stars from SLOTIS. The results are summarised in Table \ref{tab:ukirt}.

\begin{table}\begin{center}\begin{minipage}{3.25in}
      \caption{Log of spectral observations of SN~2015da}
\centering
\small
\begin{tabular}{@{}ccccc}\hline\hline
Date &Age &Tel./Instr. &Range \\
UTC &(days) & &(\AA) \\
\hline
%% Begin data
2015-02-23 &46  &Keck/DEIMOS &4850--7500 \\
2015-03-23 &74  &MMT/BCH    &5700--7000 \\
2015-04-15 &97  &LBT/MODS   &6500--8600 \\
2015-05-01 &113 &MMT/BCH    &5730--7030 \\
2015-05-11 &123 &MMT/BCH    &5710--7010 \\
2015-06-13 &156 &MMT/BCH    &5710--7010 \\
2015-07-24 &197 &Shane/Kast &3400--10800 \\
2015-09-06 &241 &Shane/Kast &3400--10.800 \\
2016-02-11 &399 &Shane/Kast &3400--10,800 \\
2016-02-16 &404 &MMT/BCH    &5700--7000 \\
2016-04-17 &465 &Bok/B\&C   &4000--8300 \\
2016-06-04 &513 &MMT/BCH    &5700--7000 \\
2016-06-08 &517 &Bok/B\&C   &4000--8000 \\
2016-06-14 &523 &Bok/B\&C   &3700--8300 \\
2016-07-14 &553 &Bok/B\&C   &3900--8100 \\
2017-01-05 &728 &Bok/B\&C   &5500--7500 \\
2017-01-07 &730 &MMT/BCH    &5720--7020 \\
2017-03-04 &786 &MMT/BCH    &5720--7020 \\
2017-03-22 &804 &MMT/BCH    &5750--7050 \\
2017-05-14 &857 &Bok/B\&C   &4500--8000 \\
2017-05-20 &863 &MMT/BCH    &5750--7050 \\
2017-05-21 &864 &MMT/BCH    &3700--9000 \\
2017-06-30 &904 &MMT/BCH    &5720--7020 \\
2017-12-03 &1060 &MMT/BCH   &5830--7130 \\
2017-12-29 &1086 &MMT/RCH   &6170--6970 \\
2017-12-29 &1086 &MMT/RCH   &8230--8980 \\
2018-04-15 &1193 &MMT/BCH   &5720--7020 \\
2018-06-18 &1257 &MMT/BCH   &5720--7020 \\
2019-01-05 &1457 &Keck/LRIS &3660--10,300 \\
2020-12-14 &2167 &MMT/BCH   &5740--7040 \\
2021-01-13 &2197 &MMT/BCH   &5700--7000 \\
2021-06-11 &2345 &MMT/BCH   &5700--7000 \\
2021-12-06 &2523 &MMT/BCH   &5700--7000 \\
2022-03-24 &2603 &LBT/MODS  &6500--8600 \\
2023-04-27 &3030 &MMT/BCH   &5700--7000 \\
%% End data
\hline \\
\end{tabular}
\label{tab:spectra}
\end{minipage}\end{center}
\end{table}

\begin{figure}
\includegraphics[width=3.2in]{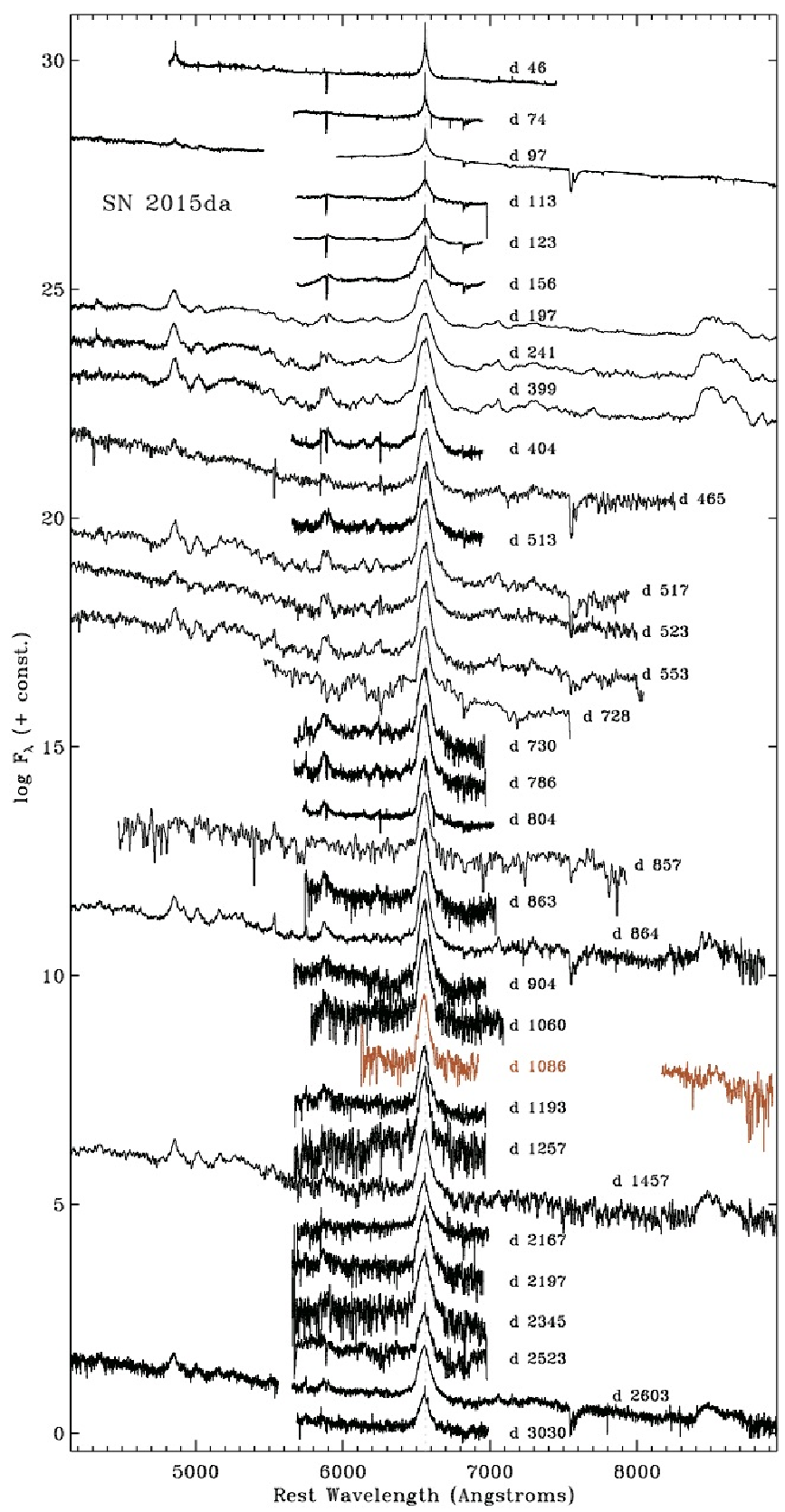}
\caption{A series of low- and moderate-resolution spectra of SN~2015da from MMT, LBT, Keck, Lick, and Bok (see Table~\ref{tab:spectra}).}
\label{fig:spectra}
\end{figure}

\subsection{Spectroscopy} \label{obs:spec}

We obtained spectra of SN~2015da using the 6.5\,m MMT with three
different instruments, including the Bluechannel (BC) spectrograph,
the Redchannel (RC) spectrograph, and BinoSpec \citep{fabricant19}. Each MMT Bluechannel
observation was taken with a 1.0 arcsec slit and either
the 1200\,lines\,mm$^{-1}$ grating covering a range of
 $\sim$5700$-$7000\,\AA, or the 300\,lines\,mm$^{-1}$ grating, covering 
 $\sim$3600$-$9000\,\AA. MMT/Redchannel observations also used the 1200\,lines\,mm$^{-1}$ grating 
 with two different tilts centred on H$\alpha$ and the Ca~{\sc ii} 
near-IR triplet.    Standard reductions were carried out using IRAF\footnote{IRAF, 
the Image Reduction and Analysis Facility (R.I.P.), is distributed by the
National Optical Astronomy Observatory, which is operated by the Association
of Universities for Research in Astronomy (AURA) under cooperative
agreement with the National Science Foundation (NSF).} including bias 
subtraction, flat-fielding, and optimal extraction
of the spectra. Flux calibration was achieved using spectrophotometric
standards observed at an airmass similar to that of each science
frame, and the resulting spectra were median combined into a single
one-dimensional (1D) spectrum for each epoch.  Late epochs of visual-wavelength spectra
obtained with Binospec on the MMT used
the 600\,lines\,mm$^{-1}$ grating centred on 6300\,\AA\ (covering a range
of $\sim$5100$-$7500\,\AA) and with a $1.0''$ slit. All data
were reduced using the Binospec pipeline \citep{kansky19},
which includes an internal flux calibration into relative flux units
from throughput measurements of spectrophotometric standard
stars.

Additional spectra were obtained using MODS on the LBT, and at the W. M. Keck Observatory 
using the Deep Imaging Multi-Object Spectrograph \citep[DEIMOS;][]{faber03} and 
the Low-Resolution Imaging Spectrometer \citep[LRIS;][]{oke95}.    We also 
obtained a few epochs of low-resolution optical spectra with the Boller \& 
Chivens (B\&C) spectrograph mounted on the 2.3\,m Bok telescope at Kitt Peak, as well as the Kast 
spectrograph \citep{ms93} on the Lick 3\,m Shane reflector.  Data reduction 
for these followed standard reduction for point sources in long-slit optical 
spectra, as above, except that the LRIS spectrum on day 1457 was reduced using the LPipe data-reduction pipeline \citep{perley19}.  All visual-wavelength spectra are corrected for 
$z=0.0067$ (rather than $z=0.0071$; see below) and a reddening of $E(B-V)=0.98$\,mag.  Our 
low/moderate-resolution spectra are plotted in Figure \ref{fig:spectra}, and details of 
the spectroscopic observations are summarised in Table \ref{tab:spectra}.

\begin{figure*}
\includegraphics[width=5.5in]{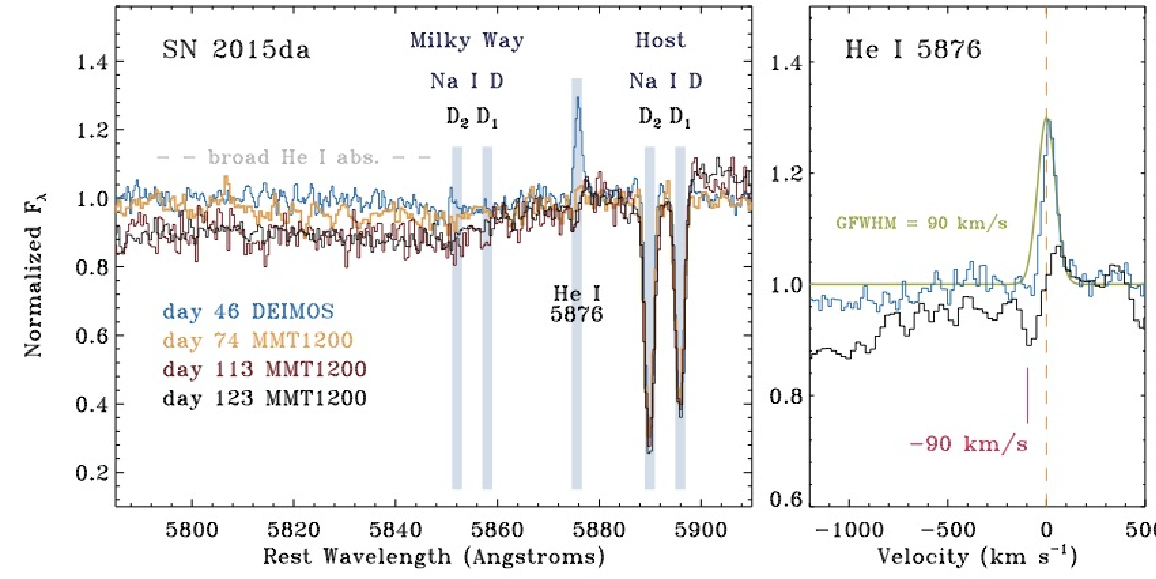}
\caption{{\it Left:} Region of the spectrum around the interstellar Na~{\sc i}~D lines as seen in our moderate-resolution spectra at early times around peak luminosity.  Wavelengths for the Na~{\sc i}~D doublet in both the Milky Way (blueshifted here) and in the host galaxy are indicated by the light-blue vertical bars, as is the wavelength of He~{\sc i} $\lambda$5876.  Note that these spectra have been corrected for a redshift of $z=0.0067$, which differs by $-120$\,km\,s$^{-1}$ from the centroid redshift of the host galaxy, presumably due to galactic rotation at SN~2015da's location. Narrow He~{\sc i} $\lambda$5876 seen on day 46 transitions to weaker emission and P~Cygni absorption at later epochs, and the very broad He~{\sc i} $\lambda$5876 absorption from the SN ejecta (in the left side of this panel) grows in strength. {\it Right:} Zoom-in on He~{\sc i} $\lambda$5876 showing only the day 46 and day 123 spectra.  The $-$90\,km\,s$^{-1}$ velocity of the P~Cygni trough on day 123 is marked in magenta, and a Gaussian with a FWHM (GFWHM) of 90\,km\,s$^{-1}$ is shown in green, centred on zero velocity.}
\label{fig:na1d} 
\end{figure*}

\begin{figure}\begin{center}
\includegraphics[width=3.7in]{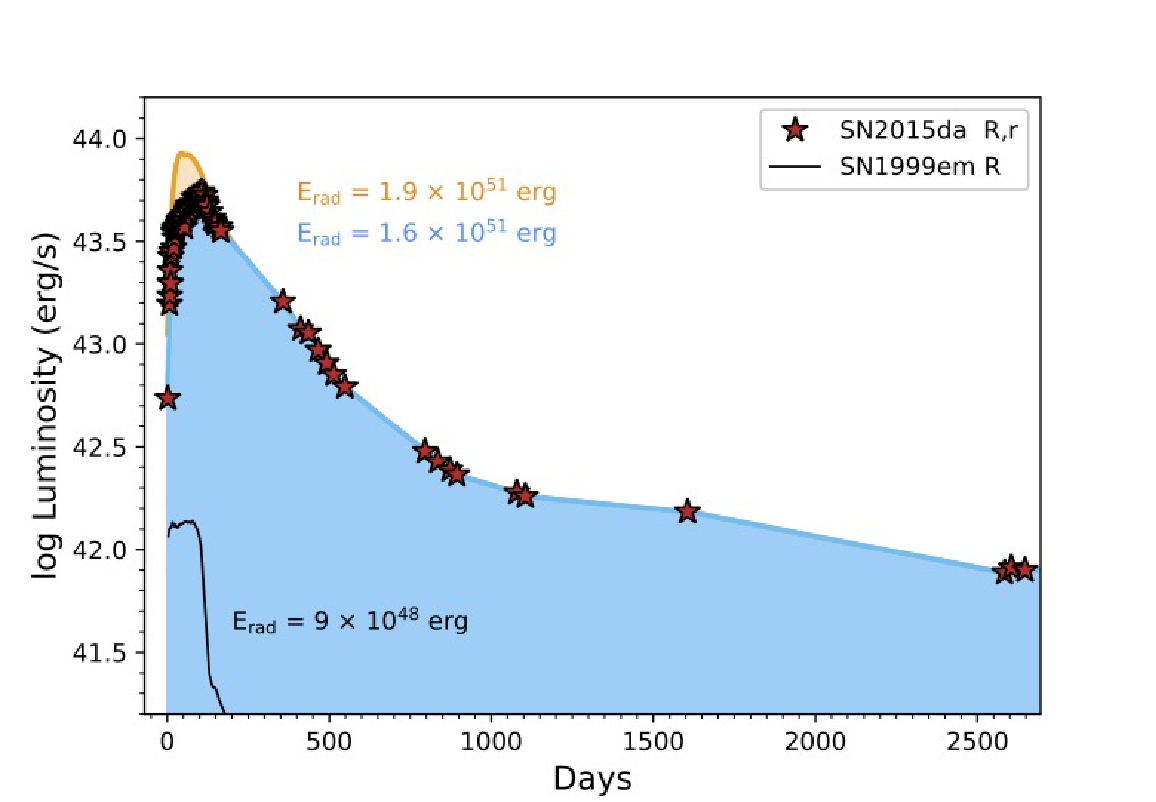}
\end{center}
\caption{The $r/R$-band light curve of SN~2015da converted to L$_{\odot}$
  values. The blue curve is interpolated between the photometry, and
  the area under the curve corresponds to a total radiated energy
  $E_{\rm rad}$ of $1.6 \times 10^{51}$\,erg with no
  bolometric correction applied.  The orange curve, which only differs
  at early times, shows an approximation of the ``pseudo-bolometric''
  light curve of SN~2015da from T20.  In this case the integrated
  radiation would be $E_{\rm rad} = 1.9 \times 10^{51}$\,erg.  The
  light curve and the much lower integrated radiation measured the
  same way are shown for SN~1999em (black), for comparison.}
\label{fig:erad}
\end{figure}

\begin{figure}\begin{center}
\includegraphics[width=2.95in]{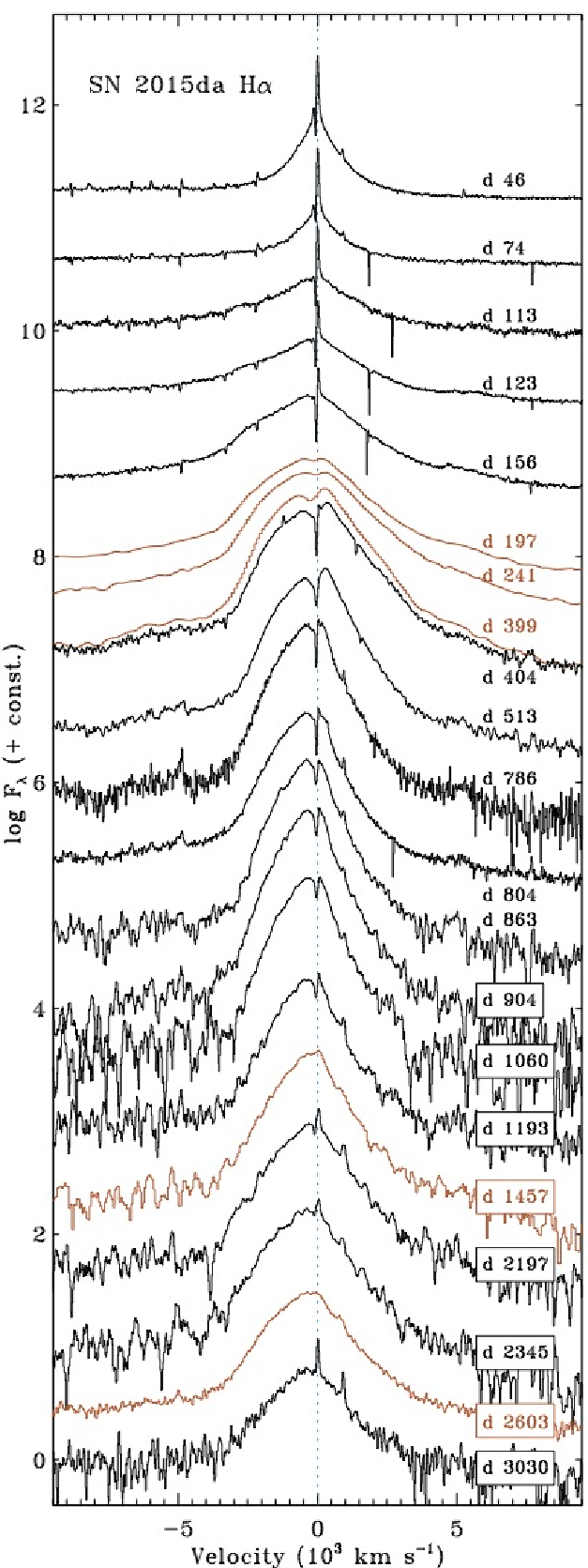}
\end{center}
\caption{Spectral evolution of the H$\alpha$ line profile as seen in a selection of spectra emphasising those with higher spectral resolution.  Some lower-resolution spectra are shown (plotted in orange) because they sample times when we lack higher-resolution spectra, or they have good signal-to-noise ratio (S/N) at times when higher-resolution spectra have relatively poor S/N. The narrow P~Cygni profiles are missing from these spectra owing to their lower resolution.  Asymmetric blueshifted profiles begin to develop after day 100 and continue until the latest observations.}
\label{fig:ha}
\end{figure}

\begin{figure}\begin{center}
\includegraphics[width=2.9in]{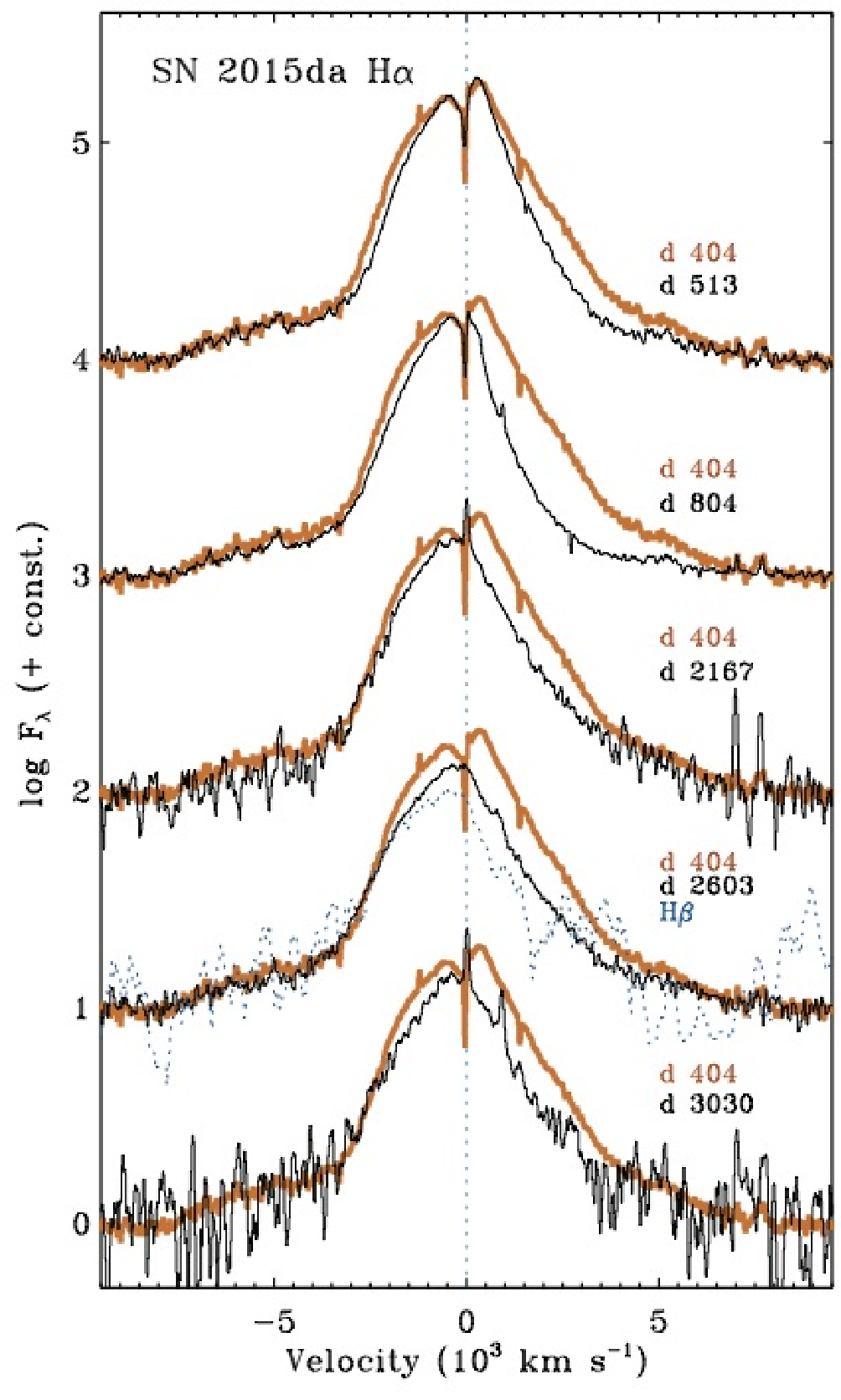}
\end{center}
\caption{A few late-time H$\alpha$ profiles from Figure~\ref{fig:ha}, where five late epochs (days 513, 804, 2167, 2603, and 3030) are shown in black, each plotted on top of the day 404 H$\alpha$ profile (orange).  For day 2603, the H$\beta$ profile is also scaled and overplotted (dashed blue) for comparison.}
\label{fig:ha.dust}
\end{figure}

\section{RESULTS}\label{results}
%\section{Analysis} \label{analysis}

\subsection{Narrow Na~{\sc i}~D, Extinction, and He~{\sc i} }

Figure~\ref{fig:na1d} shows a detail of the region of the spectrum around the interstellar Na~{\sc i}~D lines in our higher-resolution spectra obtained at early times when SN~2015da was brightest.  Strong Na~{\sc i}~D absorption is detected from the interstellar medium of the host galaxy, but Na~{\sc i}~D is not detected from the Milky Way (the spectra in Fig.~\ref{fig:na1d} are corrected for redshift, so the Milky Way components appear blueshifted as noted in the caption).  This agrees qualitatively with the low foreground reddening in the Milky Way and the relatively high reddening inferred for the host (T20).  We measure equivalent widths (EWs) of 1.3\,\AA\ and 1.1\,\AA\ ($\pm$0.05\,\AA) for the pair of lines in the host galaxy, averaged over the four spectra in Figure~\ref{fig:na1d}.  This is in reasonable agreement with the EWs measured for these lines already by T20, who noted that they are too strong to use the standard relationship for interstellar reddening \citep{2012MNRAS.426.1465P}.  We do not alter the reddening estimate, and as noted above, we adopt the value used by T20 of $E(B-V)=0.98$\,mag for the total line-of-sight reddening.

In examining this region of the spectrum, however, there are a few notable points.  First, we needed a slightly different redshift than that of the host galaxy in order to match the observed Na~{\sc i}~D lines to their expected laboratory wavelengths (shown by the light-blue bars in Fig.~\ref{fig:na1d}).  As noted in the Introduction, the host galaxy NGC 5337 has a centroid velocity that gives $z=0.0071$, but the redshift used to correct the spectra in Fig.~\ref{fig:na1d} is $z=0.0067$.  In other words, host-galaxy interstellar gas along the line of sight to SN~2015da is blueshifted by $-$120\,km\,s$^{-1}$ compared to the centroid velocity of the host, with the difference most likely due to galactic rotation at the location of SN~2015da.  This is useful when interpreting the velocities of narrow components throughout the paper.  For all remaining spectra in the paper, we correct the redshift using $z=0.0067$.

Second, this region of the spectrum reveals interesting behaviour of He~{\sc i} $\lambda$5876, and the adopted redshift correction impacts its interpretation.  We see that  He~{\sc i} $\lambda$5876 appears as a narrow emission line in our earliest spectrum (day 46), but after that the emission line weakens, and goes into absorption.  As time proceeds, we also see increasing strength of broad P~Cygni absorption, indicating that we are detecting the fast SN ejecta even at times near peak brightness.  

The right panel of Fig.~\ref{fig:na1d} shows a detail of He~{\sc i} $\lambda$5876.  Here, zero velocity is defined by the alignment of the nearby Na~{\sc i}~D lines.  Using this redshift correction, the narrow emission peak on day 46 is slightly redshifted from the rest velocity, suggesting that this is actually a P~Cygni profile, but the relatively weak and narrow P~Cygni absorption may be unresolved.  SN~2017hcc was another superluminous SN~IIn that provides a confirming example of this, where moderate-resolution spectra (comparable to the spectra here) did not show P~Cygni features, whereas dramatic narrow P~Cygni lines were revealed in high-resolution echelle spectra at the same epoch \citep{smith20}.  By day 123, it is much clearer that SN~2015da has a P~Cygni profile in He~{\sc i} $\lambda$5876, because the emission is weaker and more redshifted, and the blueshifted absorption is stronger.  The P~Cygni trough has a velocity of about $-$90\,km\,s$^{-1}$ (marked in magenta).  A Gaussian with this velocity width and centred at 0\,km\,s$^{-1}$ (shown in green) is able to match the red side of the emission on day 46, but overestimates the emission on the blue side on that date.   The Gaussian probably gives a better indication of the true intrinsic emission, while the difference between the Gaussian and the observed He~{\sc i} $\lambda$5876 emission on day 46 indicates where blueshifted P~Cygni absorption has altered the line shape.  

This is a cautionary tale that velocities measured in low- and moderate-resolution spectra, especially when adjusted to set the apparent line centre at zero, may be misleading.  In any case, this provides an indication that the mass loss shortly before explosion had an outflow speed of around 90\,km\,s$^{-1}$.  Some of the most extreme RSGs such as VY~CMa do show some knots and condensations moving this fast \citep{smith04}, but in general the outflows from RSGs are slower at $\sim$20--40\,km\,s$^{-1}$ \citep{ry98,beasor20}.  Interestingly, the recent very nearby SN~II in M101, SN~2023ixf, had narrow lines detected in high-resolution echelle spectra that indicated a similarly fast expansion speed of 115\,km\,s$^{-1}$ in the pre-SN CSM \citep{smith23}, even though the progenitor was likely a cool RSG \citep{jencson23,kilpatrick23,griffin23,ps23}.  On the other hand, an outflow speed of 90\,km\,s$^{-1}$ is on the low end of the range of quiescent wind speeds seen for LBVs, but consistent with the eruptive outflows of some LBVs \citep{smith14,smith11lbv}.  It is very similar to the slow $\sim$100\,km\,s$^{-1}$ equatorial outflow that preceded the 19th century eruption of $\eta$~Car \citep{smith18}.

% 3.2
\subsection{Light curve}

Figure~\ref{fig:lightcurve} shows the absolute-magnitude light curves
of SN~2015da compared to a few other SLSNe~IIn.  T20 have already
discussed the light curve's main peak, and we refer the reader to that
paper for detailed information.  The main information added by our
additional late-time photometry is that the unusual longevity of
SN~2015da has only continued.  At later and later times, the rate of
decline seems to slow progressively more.  From the last clear inflection
around day 1000 up to the present epoch, the decline rate is extremely
slow at only $\sim$0.00067 mag\,d$^{-1}$.  At the latest epochs,
SN~2015da still remains as luminous as the peak and plateau of a
normal SN~II-P. The sustained high luminosity throughout its evolution
points to a prolonged period of very high progenitor mass loss in the
centuries leading up to its final death; the corresponding mass-loss
rates are discussed below.

Comparing to other SLSNe~IIn, SN~2015da had a peak luminosity that was
not quite as high as that of SN~2006gy \citep{smith07}, but it lasted
much longer.  SN~2015da was like a more luminous version of the
well-studied SN~2010jl \citep{smith11,fransson14,jencson16,tsvetkov16}, and
it is perhaps most similar to SN~2003ma \citep{rest11} and SN~2016aps \citep{nicholl20,suzuki21}, which had even higher late-time luminosity.  These are the most energetic known
SLSNe~IIn.

Among these beasts, SN~2015da is unusual for its relatively slow rise
to peak.  In the first $\sim 20$\,days after explosion, SN~2015da rose
quickly, similar to the rapid initial rise times of SN~2003ma and
SN~2006gy.  However, then SN~2015da halted its rapid brightening
around day 20, and from there, it slowly crept up to its final peak
luminosity in the $r/R$ bands by day 110.  Previously, SN~2006gy had
the slowest well-documented rise among SLSNe~IIn, reaching its peak at
$\sim$70 days \citep{smith07,smith10}.  However, this arrested
brightening in SN~2015da may be somewhat misleading from only considering the
optical light curve; T20 noted that in the pseudobolometric light
curve (including near-ultraviolet and IR wavelengths), the peak
luminosity was higher and was reached by $\sim$30\,days.

Figure~\ref{fig:erad} shows the full light curve plotted in
L$_{\odot}$ values (this is corrected for distance and reddening).  Integrating the luminosity
over time provides a measure of the total radiated energy, $E_{\rm
  rad}$.  When we interpolate the observed $R/r$-band data (blue curve) and
integrate, we measure a total radiated energy (from explosion up to
day 2700) of $E_{\rm rad} = 1.6 \times 10^{51}$\,erg.  This is a
lower limit, since we have applied no bolometric correction.  The
bolometric correction will evolve with time, and is likely to be
relatively large at early times when the SN is hotter.  As
noted above, the early slow rise makes it appear as if the initial
peak of SN~2015da is missing some flux compared to the shape of the
early light curves of SN~2003ma and SN~2010jl.  Multiband photometry
indicates that SN~2015da emitted significant amounts of flux in the ultraviolet
and IR in this time, and the orange curve in Figure~\ref{fig:erad} shows
an approximation of the ``pseudo-bolometric'' light curve presented
by T20.  This curve has an integrated radiated energy of $E_{\rm rad}
= 1.9 \times 10^{51}$\,erg.

Observations indicate substantial IR flux after day 400 (T20), and
there is also likely to be significant radiation in X-rays at late
times after $\sim 1000$\,days when the optical depth is low enough for
the X-rays to escape.  In analysing the similar light curve of
SN~2003ma, \citet{rest11} measured $E_{\rm rad} = 9 \times 10^{50}$\,erg, and with a phase-dependent bolometric correction, estimated a
total bolometric radiated energy of $4 \times 10^{51}$\,erg.  
Similarly, \citet{nicholl20} estimated $E_{\rm rad} = 5 \times 10^{51}$\,erg  for SN~2016aps.  
It is unclear if precisely the same correction should be applied to SN~2015da, but
nevertheless, it is likely that SN~2015da's true bolometric radiated
energy should be at least $3 \times 10^{51}$\,erg.  This is comparable,
within the uncertainties, to the inferred values for SN~2003ma and SN~2016aps, and together these
values are more than those of any other known SNe.  A typical SN~II-P, like
SN~1999em shown in Figure~\ref{fig:erad}, radiates $\sim$200 times
less energy.  Since $E_{\rm rad}$ is only a portion of the total
energy budget (ignoring neutrinos, there is still the kinetic
energy of the final coasting speed of the swept-up CSM shell and the kinetic energy of the unshocked SN ejecta; see
below), this indicates a rather energetic explosion mechanism
exceeding 1 FOE by a significant margin.  This, combined with the mass
budget discussed later, clearly rules out SNe~Ia in dense CSM for
these SLSN~IIn events.

% 3.3
\subsection{Spectral evolution}

Figure~\ref{fig:spectra} shows our full series of optical spectra of
SN~2015da.  T20 already discussed the overall behaviour of the spectrum
and comparison to other events, so this discussion here will be brief.
We do note a few differences in the implications from our data, however.

In general, SN~2015da displays the classic optical spectral evolution
observed in many SNe~IIn \citep[see][]{smith08tf}.  At early phases up to
and around the time of peak luminosity (roughly the first 100--150\,days), 
SN~2015da shows a smooth blue continuum with strong, narrow
Balmer emission lines that have smooth Lorentzian-shaped line wings.
This is usually attributed to a phase where the photosphere is
actually in the CSM ahead of the forward shock, and the line profiles
arise from narrow emission from pre-shock gas, but with line wings that are broadened
by electron scattering \citep{chugai77,smith08tf}.  While SN~2015da
fades from its peak, the spectrum transitions as is typical of
SNe~IIn.  The Lorentzian wings of Balmer lines morph into more
irregular, multicomponent, and asymmetric line-profile shapes.  The
smooth blue continuum begins to give way to a pseudo-continuum of many
blended emission, absorption, or P~Cygni lines in the blue, and it
shows broad and intermediate-width emission features from the SN
ejecta and shocked gas like the Ca~{\sc ii} near-IR triplet and He~{\sc i}
emission lines.

A key point is that during this decline after peak (days 100--1000),
the spectra exhibit clear evidence of emission and absorption from the
freely expanding SN ejecta, in addition to the dense post-shock shell.
This is seen in the broad Ca~{\sc ii} near-IR triplet (which exhibits no narrow or intermediate-width components), the broad wings of
H$\alpha$ (see below), and the broad P~Cygni absorption in He~{\sc i}
(see Fig.~\ref{fig:na1d}).  We infer that at these times, the
photosphere can no longer be ahead of the forward shock in the CSM.
This, in turn, is consistent with the disappearance of the Lorentzian
wings of Balmer lines that were due to electron scattering, allowing
us to see the kinematically broadened line profiles from post-shock
gas and ejecta.  Such a distinction is important, because we can then
use the width of Balmer lines (after this transition occurs at roughly
day 110--120) to trace the expansion speed of the shock running through CSM
(see below).

At late times after day 1000, the spectrum evolves very slowly.  The
emission features from the SN ejecta, like the Ca~II near-IR triplet, actually
fade somewhat, and the spectrum is dominated by broad or
intermediate-width H$\alpha$ with a relatively weak continuum.
H$\alpha$ is the brightest line in the optical spectrum at all epochs,
and its detailed evolution is discussed next.

% 3.4
\subsection{H$\alpha$ line-profile evolution}

Figure~\ref{fig:ha} shows the evolution of the H$\alpha$ profile,
zooming in on the velocity range of $\pm 9500$\,km\,s$^{-1}$.  The most
salient changes with time are in the shape of the intermediate-width
component, the strength of broad emission wings, and the evolution (or
lack thereof) in the narrow H$\alpha$ emission and P~Cygni absorption.

As noted above, H$\alpha$ shows the typical evolution seen
in SNe~IIn, from narrow emission with symmetric Lorentzian wings at early
times, to a more complex, asymmetric, multicomponent profile at later
times.  This transition occurs at different times in each
SNe~IIn; in SN~2015da, the transition occurs gradually around the
time of peak $R$-band luminosity, which is around day 110.  The first
two epochs (days 46 and 74) exhibit symmetric Lorentzian wings.  Day
113 still appears vaguely Lorentzian in shape, but the wings are
already showing asymmetry, with excess flux on the blueshifted side
(this is discussed more below).  By day 123, the profile is clearly
not Lorentzian, being strongly skewed in shape and having a clear blue
excess.  From that
point until the end of our spectral sequence, all of the H$\alpha$
profiles have an asymmetric, blueshifted profile caused by a deficit
of flux on the red wing of the intermediate-width component. This is
discussed more in the next subsection.

Midway through the decline from the main peak of the light curve, the
H$\alpha$ profile has developed not only a net blueshift in its
flux-weighted centroid, but a profile that is obviously asymmetric and
skewed (i.e., a different shape in the blue and red wings of the line).
The day 404 profile in Figure~\ref{fig:ha}, observed with relatively
high resolution and good S/N, demonstrates this clearly.
The blue wing of the intermediate-width component is rounded, with a
relatively abrupt transition to the broader wing at roughly $-$3000\,km\,s$^{-1}$.  There is a net blueshift to the line centroid, with a
larger half width at half-maximum intensity (HWHM) on the blue side (1660\,km\,s$^{-1}$) than on the red side (1290\,km\,s$^{-1}$).  Despite the overall
blueshift, the peak of the emission is actually on the red side of the
line at $+$350\,km\,s$^{-1}$.  From that peak, there is a steep and
roughly linear (at least as it appears in the log scale in
Fig.~\ref{fig:ha}) drop in flux on the red wing, which then more
gradually blends into the broader component at $+$5000 to $+$7000\,km\,s$^{-1}$.  There is no way to match this shape with any symmetric
Lorentzian profile, even if that Lorentzian has a blueshifted
centre.

From day 404 onward, the H$\alpha$ profile becomes even more
asymmetric.  Figure~\ref{fig:ha.dust} compares the day 404 profile
shape (in orange) to several subsequent epochs (days 513, 804, 2167, 2603, and 3030) with good S/N and resolution (in black).  In the
blue wing, there are slight differences in the HWHM velocity, but in
the red wing the differences are dramatic, with all epochs after day
404 showing a strong deficit of flux on the red side (even though the
day 404 profile is itself already asymmetric).  The missing flux in
the red wing is greatest in the day 804 profile, and then this
difference lessens as later epochs after day 2000 resemble the day 513
deficit.

After the time of peak brightness, the broad wings of H$\alpha$ do not change
dramatically, and appear relatively symmetric about zero velocity.
This suggests that whatever is causing the deficit of flux in the
red side of the intermediate-width component is not blocking as
much of the receding emission from SN ejecta.  This is relevant to
the interpretation discussed in Section 4, because it means that the
blocking agent is primarily in the post-shock CDS, not in the central SN
ejecta.  The evolution of the narrow emission and absorption component
is discussed in a later subsection, after we explore functional fits
to the intermediate and broad components of H$\alpha$.

\begin{figure}\begin{center}
\includegraphics[width=2.9in]{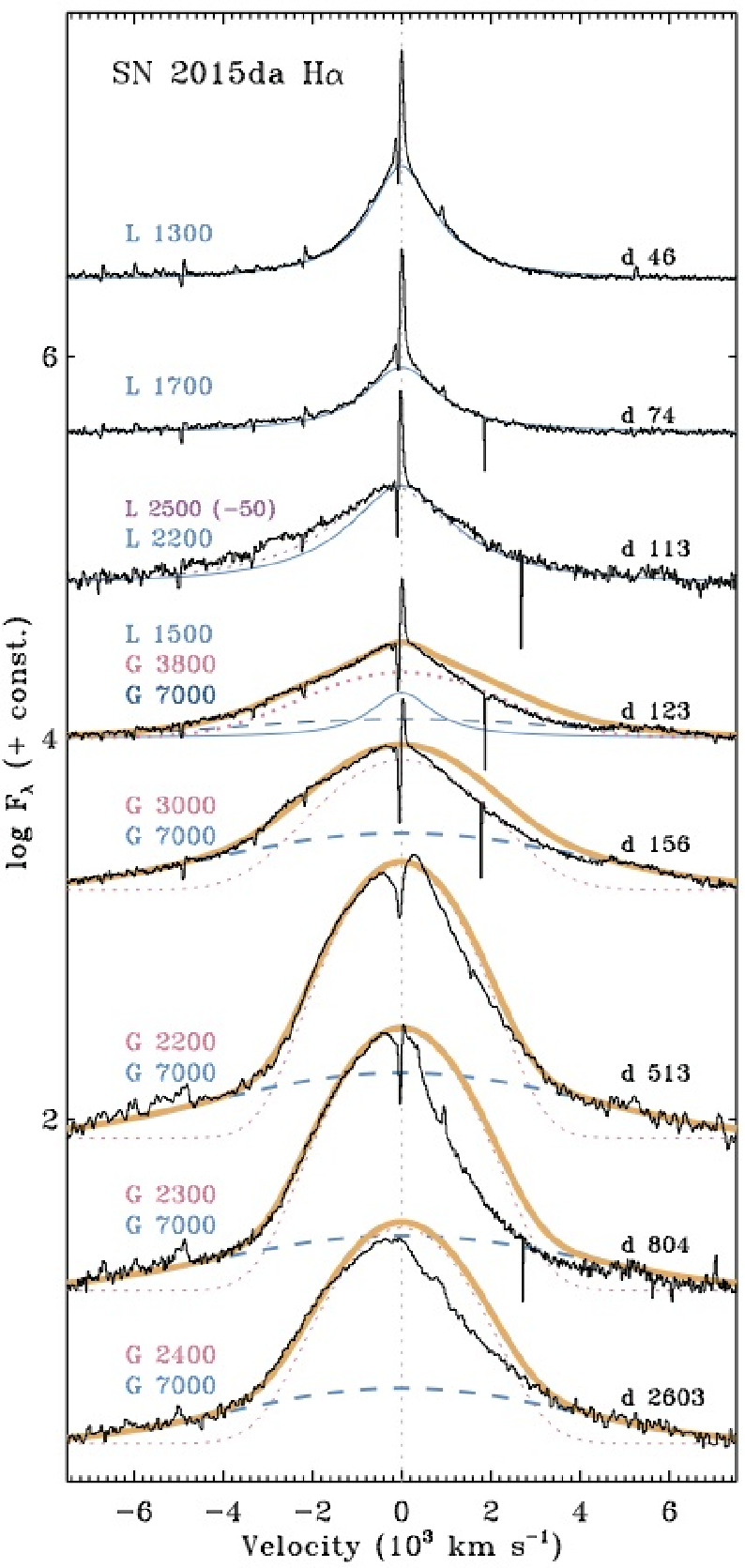}
\end{center}
\caption{A series of observed spectra showing the H$\alpha$ line
  profile in SN~2015da (black), compared with smooth Lorentzian or
  Gaussian profile shapes (blue or magenta), and the sum of these
  (thick orange curve) when more than one is used.  For day
  113 we also plot a FWHM = 2500\,km\,s$^{-1}$ Lorentzian that is
  shifted by $-$50\,km\,s$^{-1}$ (purple dashed line).  Each epoch lists the
  Lorentzian (L) or Gaussian (G) value(s) for the FWHM in km\,s$^{-1}$ to
  the left of the line.  At later epochs, these curves intentionally
  ignore the red wing of the line, presuming that there is a deficit
  of flux in the red wings due to internal obscuration (see text).}
\label{fig:ha.fits}
\end{figure}

% 3.5
\subsection{H$\alpha$ profile fits and velocity}

Here we explore functional approximations to the H$\alpha$ line shape
and its evolution with time.  Figure~\ref{fig:ha.fits} shows a
selection of observed H$\alpha$ profiles in our spectra (black)
compared to Lorentzian or Gaussian functional shapes (or combinations
of them).

As noted above, the first two epochs on days 46 and 74 have wings that
are well matched by the shape of a symmetric Lorentzian function
centred on zero velocity, while the narrow emission sits atop this
Lorentzian.  This is typical of SNe~IIn at early times.  Physically,
this line shape is thought to arise when narrow-line photons from the
slow pre-shock CSM must escape from high optical depths, and where thermal
electron scattering broadens the profiles to the red and blue by the
same amount \citep{chugai77,smith08tf}.  Lorentzian FWHM values of
1300 and 1700\,km \,s$^{-1}$ on days 46 and 74 (respectively) are typical
of early SLSNe~IIn spectra
\citep{fransson14,smith08tf,smith10,smith20,dickinson23}. These
Lorentzian shapes topped by narrow P~Cygni emisson/absorption are
reproduced in radiative-transfer simulations \citep{dessart15}.

Looking closely, a Lorentzian profile matches the line wings on day 46
in great detail, within the limits of S/N.  By day 74,
however, we can already see a small discrepancy between the observed
line wings and the Lorentzian, with a slight excess of observed flux
in the blue wing at $-$2000 to $-$4000\,km\,s$^{-1}$.  This excess is
relatively small, but more than the noise.  By day 113, right around
the time of peak luminosity, a Lorentzian function centred at zero
utterly fails to capture the shape of the observed line wings.  A 2200\,km\,s$^{-1}$ Lorentzian centred on the narrow emission on day 113 matches the red wing, but does
not come close to matching the blue wing.  Even a broader
2500\,km\,s$^{-1}$ Lorentzian that has its centroid artificially shifted
by $-$50\,km\,s$^{-1}$ from the narrow emission (purple dashed  line in
Fig.~\ref{fig:ha.fits}) still underestimates the blue wing on day 113,
although it comes closer than a centred Lorentzian.  This discrepancy
with Lorentzian profiles only worsens as time proceeds after day 113.
For all epochs after day 113, we therefore use a combination of
multiple Gaussian components to match the line shape.

Here our analysis differs significantly from that of
SN~2015da's spectra by T20.  T20 interpreted the blueshifted H$\alpha$
emission profiles the same way that \citet{fransson14} interpreted
the similar blueshifted profiles in SN~2010jl --- as being due
to broadening by electron scattering at all epochs, but where a
symmetric Lorentzian function has its centroid shifted to the blue
relative to the narrow emission at zero velocity.  As noted by
\citet{smith20}, the physical basis for this idea is doubtful, since
the Lorentzian shape of the electron-scattering wings must be
symmetric about the source of the original narrow-line photons that
get broadened by electron scattering.  However, in this case of SN~2015da, and
previous examples of SNe~IIn that show a similar blueshift in their
intermediate-width components, the narrow lines are {\it not}
blueshifted by the same amount as the centroid of the putative
Lorentzian function.  Simulations also show line wings that are
symmetric about zero velocity (i.e., the centroid of the narrow
emission) until late times (around day 200 in the case of SN~2010jl)
when the wings are influenced by the post-shock gas that becomes
visible \citep{dessart15}.  Here we therefore do not adopt this same
unphysical picture of symmetric but artificially blueshifted
Lorentzians.

Instead, a more plausible explanation for the asymmetry is that the
line profiles appear asymmetric and blueshifted because some portion of
the receding (redshifted) emission is blocked from our view.  As noted
by \citet{smith12} for the case of SN~2010jl, this may occur either
through occultation by the SN
ejecta photosphere (at early times) or by dust grains
that have formed in the SN ejecta or shocked CSM (at
later times).

The dust vs.\ electron-scattering interpretations of the blueshift are
discussed more in Section 4; we would normally not mention the
physical interpretation in this stage of the analysis, but in this case,
the physical picture influences how we choose to measure velocities
from the observed spectra.  One specific quantity of prime interest
that we wish to measure is the velocity of the forward shock
running into the CSM.  In T20's approach, the lines are broadened
by electron scattering, and so the observed line widths
provide no information about the speed of the forward shock, because
the width is not a kinematic width.  In the interpretation where the
intermediate-width emission arises from post-shock gas, but appears
asymmetric because the far side is blocked by the photosphere or dust,
 accounting  for this fact is important. If we want to derive
the speed of the forward shock (which in SNe~IIn is the same as the
speed of the cold dense shell (CDS) that emits the intermediate-width component), then we
wish to know the intrinsic FWHM of the line.  However, if the
redshifted emission is selectively blocked from our view, then the
observed FWHM will clearly underestimate the true intrinsic
speed. Instead, the HWHM of the blue side of the line is more
indicative of the intrinsic width of the line before extinction shapes
the line, because the approaching material is not blocked.
In terms of simple functional approximations of the line shape,
this is equivalent to fitting the blue wing of the line, and allowing
the observed flux to fall below these curves on the red side or even
at line centre.

We adopt this approach for functional fits to all spectra taken
after the time of peak luminosity.  Examples of the late-time spectra
with multicomponent Gaussians are shown in Figure~\ref{fig:ha.fits}.

From day 123 onward, we could not find any Lorentzian (even with a
shifted centre) that provided an acceptable match to the line shapes
(although on day 123, there may still be some weak emission from a
Lorentzian profile that contributes to the total line profile on top
of the broad and intermediate-width Gaussians).  After peak brightness, all
epochs require at least two Gaussians to match the blue side of the
profile.  Because we suspect that dust blocks the red side, we used
Gaussians that are centred on zero velocity, but we ignore the red
wing.  On the red side of the line, the difference between the
functional profiles and the observed red wing clearly demonstrates the
asymmetry, and indicates how much line flux is missing.

For each of the late-time spectra (day 123 onward), we found that a
broad FWHM = 7000\,km\,s$^{-1}$ Gaussian gave an adequate approximation
of the broad wings, presumably due to emission from the fast SN
ejecta.  Note that well after day 100, the SN ejecta are no longer
significantly heated by internal radioactivity.  Instead, the fast
emitting SN ejecta are probably heated as they approach the reverse
shock by backwarming \citep[see][]{smith08tf}.  In principle, this speed should decline over
time, as the speed of ejecta that reach the reverse shock must
decrease with time.  However, the S/N of the fainter wings
is not sufficient to accurately constrain this reduction in speed, so
we use the same Gaussian FWHM for all epochs.  This is not an accurate
sampling of the SN ejecta speed, but is needed to provide a broad
base for fitting the intermediate-width components.

Importantly, however, the observed broad wings are not asymmetric.  We
matched the broad 7000\,km\,s$^{-1}$ Gaussian curves to the blue wings,
but they also match the observed flux on the red wing beyond
velocities of about $+$4000 or $+$5000\,km\,s$^{-1}$.  This is
important, because it indicates that at later epochs, the emission
from the receding fast SN ejecta does not suffer the same obscuring effects as
the emission from the post-shock CDS that emits the intermediate-width
component.  This is an important clue to the location of the obscuring
material, and we return to it later.

\begin{figure}\begin{center}
\includegraphics[width=3.5in]{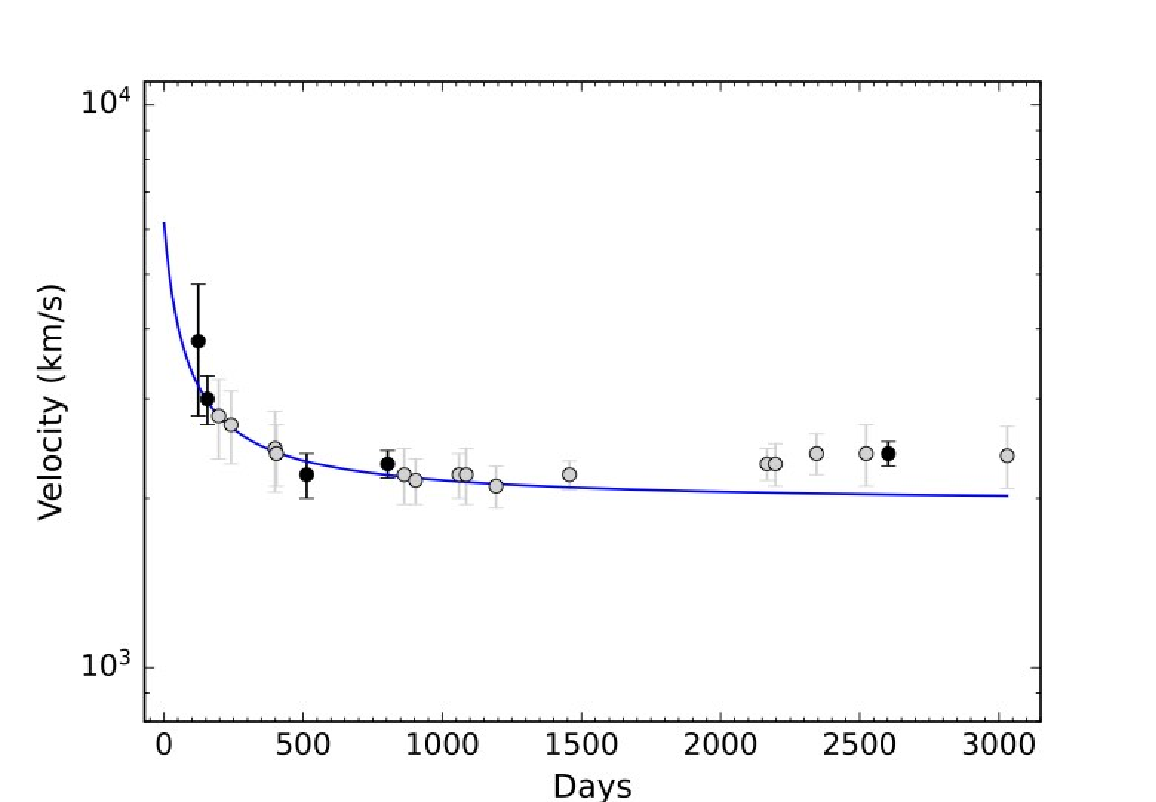}
\end{center}
\caption{The speed of the forward shock as a function of time inferred
  from spectra.  Measurements correspond to the FWHM of the
  intermediate-width Gaussian components in Figure~\ref{fig:ha.fits},
  which are only matched to the blue sides of the lines (these are
  equivalent to 2 times the HWHM values for the blue wings).  The blue curve
  is a smoothly dropping velocity that we use to approximate the CDS
  velocity later in our analysis, and this curve intentionally ignores the rise 
  in apparent velocity after day 2000 (see text).}
\label{fig:vel}
\end{figure}

% 3.6
\subsection{Speed of the forward shock}

After the time of peak brightness, the intermediate-width components are tracing emission from the CDS, and therefore can provide an estimate of the important value
for the speed of the forward shock running into the CSM.
Figure~\ref{fig:ha.fits} shows how we estimated these FWHM values for
several representative epochs (these correspond to the Gaussian FWHM
values listed in magenta and shown by a dotted magenta curve in the
plot).   These FWHM values are approximate, since we are
matching the profile shape mainly to the blue wing, as the red wing
may be missing due to extinction.  Uncertainties in these FWHM values
are typically 100--300\,km\,s$^{-1}$, depending on the data quality, the
irregularity of the shape, and in some cases the uncertain overlap
with the broad component.

Although Fig.~\ref{fig:ha.fits} shows only a few examples, we
measured this intermediate-component FWHM in all of our spectra.
Fig.~\ref{fig:vel} plots the resulting FWHM velocities of the
intermediate-width components in our spectra from day 156 onward,
after the intermediate-width profiles are no longer dominated by
Lorentzian profiles.  The blue curve in Fig.~\ref{fig:vel} is a
simple smoothly declining velocity passing through these data,
approximating the speed of the CDS that decelerates as it gets
mass-loaded with CSM.  This is used later in our analysis.

Curiously, Fig.~\ref{fig:vel} shows that the line width increases
slightly again after day 1500, rising from a minimum of 2100\,km\,s$^{-1}$ around day 1200 up to 2400\,km\,s$^{-1}$ in our last epoch
around day 3030.  There are two potential explanations for this, but it is difficult to determine which is correct.
First, it could mean that the speed of the CDS is actually
accelerating.  By $\sim$1200 days, SN~2015da may have completely outrun
its dense shell of CSM.  Continuing to expand into a rarefied medium,
the CDS would still be pushed by the fast SN ejecta hitting the
reverse shock, so the shock might accelerate outward through the dropping
density gradient.  SN ejecta hitting the reverse shock still have a speed of 7000\,km\,s$^{-1}$
at that time.  A second possible explanation is observational; as
the intermediate-width component gets fainter, it becomes somewhat
blended with emission from the SN ejecta and reverse shock, and
 the contribution of some broader emission may widen the intermediate-width
component.  This is the first time to our knowledge
that the intermediate-width emission component has been seen to
broaden at late times.  In any case, the blue curve in Fig.~\ref{fig:vel} ignores this late increase.

\begin{figure}\begin{center}
\includegraphics[width=2.8in]{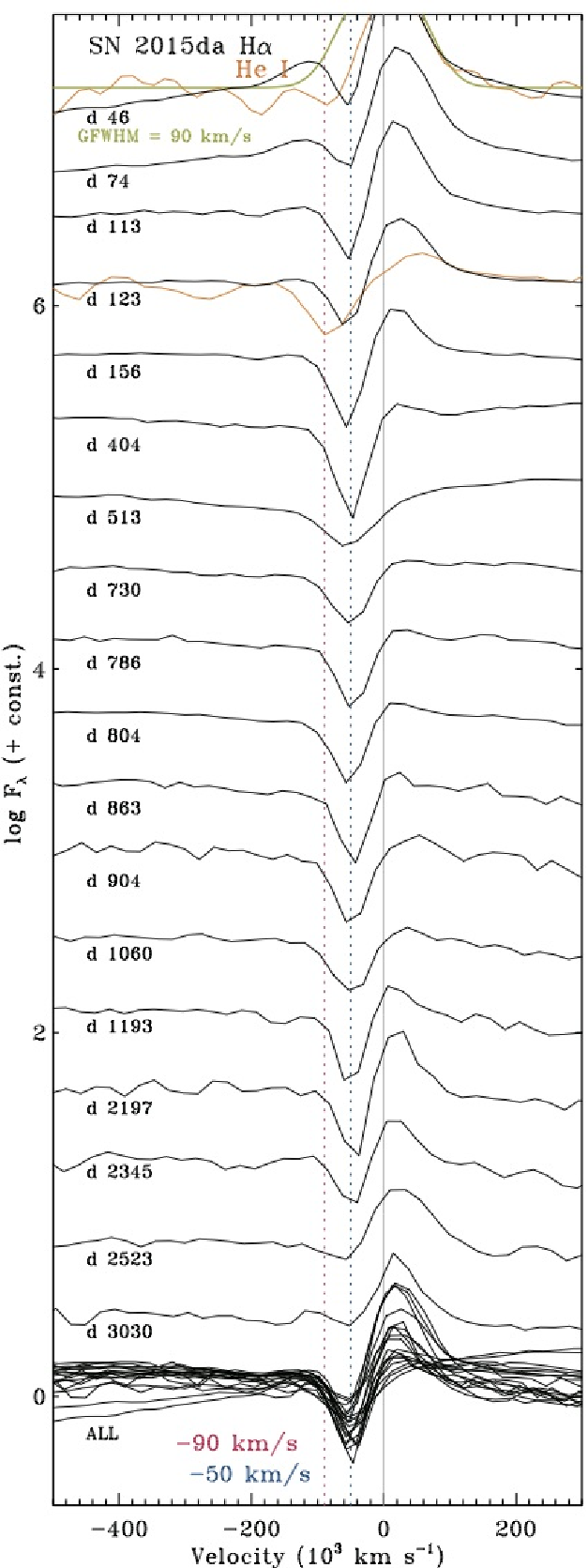}
\end{center}
\caption{Evolution of the line profile for the narrow P~Cygni
  component of H$\alpha$ (black), tracing the expansion of the
  unshocked CSM.  For days 46 and 123, we show the same profiles of
  He~{\sc i} $\lambda$5876 (in orange) that were plotted in the right
  panel of Figure~\ref{fig:na1d}, for comparison, as well as the same
  FWHM = 90\,km\,s$^{-1}$ from that figure (in lime green).  The
  vertical dotted lines in magenta and blue show reference values of
  $-$90\,km\,s$^{-1}$ and $-$50\,km\,s$^{-1}$, respectively.  $-$50\,km\,s$^{-1}$ roughly matches the trough of the P~Cygni absorption of
  H$\alpha$, whereas $-$90\,km\,s$^{-1}$ approximately matches the trough of
  the P~Cygni absorption of He~{\sc i} and the blue edge of the P~Cygni absorption in H$\alpha$.  At the bottom, all the H$\alpha$
  profiles in the figure are overplotted together with no vertical
  offset.}
\label{fig:narrow}
\end{figure}

% 3.7
\subsection{Narrow components from CSM}

In SNe~IIn, the namesake narrow emission components provide important
information about the pre-shock CSM, and hence, valuable information
about the pre-SN state of the progenitor star.  Above (see
Fig.~\ref{fig:na1d}), we already mentioned that 
He~{\sc i} $\lambda$5876 shows a narrow emission component and
a P~Cygni absorption component at $-$90 km s$^{-1}$.  That He~{\sc i}
line is clearly detected in only two of our spectra (days 46 and 123).
Here we consider the narrow component of H$\alpha$ and its evolution
over time.  Figure~\ref{fig:narrow} shows a sequence of spectra that
focuses on the narrow component of H$\alpha$ as seen in some of our
spectra with relatively high resolution and good S/N, and the two epochs where He~{\sc i} is detected are overplotted
in orange.

From Figure~\ref{fig:narrow}, it is clear that while the strength of
narrow H$\alpha$ emission and P~Cygni absorption fluctuates, the width
of the emission and the velocity of the P~Cygni absorption are
remarkably steady.  The minimum of the P~Cygni absorption trough is at
roughly $-$50\,km\,s$^{-1}$ (vertical blue dotted line) from our first
epoch to the last.  The blue edge of the P~Cygni absorption is at
roughly $-$90\,km\,s$^{-1}$ (vertical magenta dotted line), although
this blue-edge velocity does seem to fluctuate more than the velocity
of the absorption minimum, perhaps in part because the underlying
intermediate-width emission component is varying as well.  In the
first spectrum on day 46, when the narrow emission component is
strongest, we show the same FWHM = 90\,km\,s$^{-1}$ Gaussian profile
from Fig.~\ref{fig:na1d} (in lime green).  That Gaussian was found to
capture the emission profile of the He~{\sc i} line quite well
(accounting for the fact that the blue side of the line is absorbed),
and this appears to match the width of H$\alpha$.  From
this, and the blue edge of the H$\alpha$ P~Cygni absorption, we
therefore adopt a constant CSM expansion speed of 90\,km\,s$^{-1}$ in
our analysis below.

A caveat is in order, since this inferred expansion speed of 90\,km\,s$^{-1}$ is dangerously close to the resolution at H$\alpha$ in these
spectra, which is $\sim$70\,km\,s$^{-1}$.  These narrow profiles are
therefore underresolved.  When higher-resolution echelle spectra are
available for other SNe~IIn, as in the case of SN~2017hcc
\citep{smith20}, we find that our 1200\,lines\,mm$^{-1}$ grating spectra with
MMT/Bluechannel do miss some important information about the line
shape.  The blue edge and transition from absorption to emission are
likely sharper than shown in these data, so an uncertainty of 15--20\,km\,s$^{-1}$ is likely for the inferred velocity.
Nevertheless, the spectra in Figure~\ref{fig:narrow} do {\it not} show a
systematic gradual increase or decrease in the velocity of the P~Cygni
absorption, so the conjecture of a relatively constant expansion
velocity in the CSM ahead of the shock is probably correct.

A constant CSM speed has important physical implications.  SN~2015da was
extremely luminous, and it sustained a high luminosity for an unusually
long time compared to most SLSNe~IIn, requiring a relatively large
radial extent in the dense CSM.  A constant CSM expansion speed
suggests that the very strong pre-SN mass loss was in the form of a
very {\it strong but sustained steady wind} or a long series of many similar episodic events, rather than a single
short-duration burst of mass loss.  A sudden pre-SN outburst or
explosion would likely lead to a Hubble-like flow in the CSM, where
expansion speed is proportional to distance from the star, translating
to an increase in the observed outflow speed in later spectra. A clear sign of a
relatively sudden burst of mass loss 8\,yr before the SN was seen in
similar data for SN~2006gy \citep{smith10}, so SN~2015da offers an
interesting counterpoint showing a constant-speed wind.

The $\sim$90\,km\,s$^{-1}$ outflow speed is seen in all our spectra until
at least 3000\,days after explosion.  This provides an important
constraint on the minimum duration of pre-SN mass loss, assuming
constant expansion speed.  The time it took for the progenitor to
create the dense CSM is $t_{\rm CSM} = t_{\rm SN} \times V_{\rm SN}
/ V_{\rm CSM}$, where $t_{\rm SN}$ is the time of observation since
the SN exploded, $V_{\rm SN}$ is the speed of the forward shock, and 
$V_{\rm CSM}$ is the expansion speed of the CSM.  With a forward shock
expanding at roughly 2100\,km\,s$^{-1}$ (this is the minimum speed; see
above), this translates to a strong pre-SN wind that lasted for at
least 160\,yr before core collapse.  Given the higher $V_{\rm SN}$ speeds during
the earlier parts of the light curve (Fig.~\ref{fig:vel}), 200\,yr is a better
estimate.  This length of time helps to clearly rule out some
mechanisms, like wave driving, as the culprit for the pre-SN mass
loss, since wave driving can only operate effectively on timescales of
around 1\,yr before core collapse, as noted in the Introduction.  Some
other mechanism(s) must be responsible for the extreme mass loss by
SN~2015da's progenitor. 

Finally, it is interesting to note that while the widths of emission
components agree for H$\alpha$ and He~{\sc i}, the He~{\sc i} P~Cygni
absorption seems to indicate higher expansion speeds than H$\alpha$.
This is most clear on day 123 in Figure~\ref{fig:narrow}, where the P~Cygni minima of He~{\sc i} and H$\alpha$ are at $-$90 and $-$50\,km\,s$^{-1}$ (respectively), and the blue edge of He~{\sc i} absorption extends
to $-$160\,km\,s$^{-1}$, almost double that of H$\alpha$.

Curiously, the situation is exactly the opposite of what was seen
recently in SN~2017hcc, which is another SLSN~IIn.  In SN~2017hcc, the
widths of emission components agreed for H and He lines, as is the
case here, but the disagreement in absorption velocities was flipped:
in SN~2017hcc, the H Balmer lines showed faster
velocities in the P~Cygni absorption, and 
the He~{\sc i} P~Cygni absorption had slower expansion velocities \citep{smith20}.  This was
attributed to geometric effects, where the velocity and excitation
level are latitude dependent when an asymmetric disk-like CSM geometry
is hit by the SN.  Since SN~2015da also has asymmetric CSM (see
below), and since this effect depends on viewing angle, this could
signify that SN~2015da is viewed from a substantially different
latitude than SN~2017hcc.  The fact that the H/He velocity
disagreement is flipped in these two SLSNe argues for a geometric
viewing-angle effect rather than an alternative explanation where
pre-shock acceleration has a stronger influence on higher-ionisation
lines.  If the correct explanation were the latter, we should expect
the physics to work the same way in both objects.

%% sect 4
\section{Discussion}

%% 4.1
\subsection{Asymmetry}

SN~2015da is remarkable both in its high peak luminosity that makes it
an SLSN~IIn, and in how it sustained such a high luminosity over time.
This is even more remarkable when we consider evidence that the CSM is
asymmetric, since asymmetry tends to reduce the global efficiency of
converting kinetic energy into light, simply
because asymmetric CSM intercepts only a fraction of the available ejecta.

Earlier we noted a few clues that the CSM interaction in SN~2015da is
likely to be asymmetric.  The most important clue is that we see the
fast SN ejecta directly, even at early times.  We see very broad
emission wings in the H$\alpha$ emission-line profiles; moreover,
low-resolution spectra reveal broad emission from the Ca~{\sc ii} near-IR
triplet that is usually attributed to the fast SN ejecta, and in this case shows no intermediate-width or narrow emission components at these early times.  These are
seen around the time of peak luminosity and afterward.  In addition,
very broad P~Cygni absorption is detected in He~{\sc i} $\lambda$5876
out to $-$10,000\,km\,s$^{-1}$.  A hint of this broad absorption
is already present in the day 46 spectrum, and is clearly present by
day 74 and afterward (see Fig.~\ref{fig:na1d}).  Thus, we are
already seeing direct emission and absorption from the fast, freely
expanding SN ejecta, before the time of peak $R$-band luminosity.  This is
not possible in a spherical model for an SLSN, where the photosphere is
still ahead of the shock in the pre-shock CSM at these phases
\citep{dessart15,smith08tf,woosley07}.  Seeing the SN ejecta directly
at such an early epoch requires that along some lines of sight to the
SN ejecta, the optical depth is lower than required for other
directions through the opaque CSM interaction region.  This requires
asymmetric CSM.

In fact, clear evidence for significant asymmetry in SN~2015da is
provided by spectropolarimetry.  \citet{bilinski23} presented
multi-epoch spectropolarimetry of SN~2015da taken at early epochs up to
and around the time of the main luminosity peak.  During this phase
SN~2015da showed a continuum polarization of $\sim 3$\%, which is on
the high side for SNe~IIn and indicates a significant degree of
asymmetry.

If the CSM of SN~2015da is significantly asymmetric, then this is
important to keep in mind when evaluating the global energy budget of
the event discussed in the next section.  When CSM interaction is the
dominant power source for a bright SN, some fraction ($f$) of the
available kinetic energy of ejecta reaching the shock (i.e., the explosion kinetic energy, $E_{\rm exp}$) is converted to
light, and so the integrated radiated energy that we observe  ($E_{\rm
  rad}$) is obviously only a lower limit to the explosion energy:
 $E_{\rm exp} \ge E_{\rm rad}/f$.  For simulations of SLSNe with spherical CSM
interaction, $f$ can be quite high (i.e., 0.3 to 0.5,
for example; \citealt{vm10}).  Now suppose that the CSM is asymmetric,
and that the CSM intercepts only some fraction $q$ of the $4\pi$
steradians of the explosion.  For a thick disk or the waist of a
bipolar CSM shell, $q$ might plausibly be 0.1--0.4.  If the asymmetric
CSM can only tap into a fraction of the available kinetic energy, then
some of the ejecta expand without experiencing strong interaction;
thus, $E_{\rm exp} \ge E_{\rm rad}/fq$.  With both correction factors
$f$ and $q$, if the CSM is asymmetric, we can easily have $E_{\rm exp}
\approx (5-10) \times E_{\rm rad}$ (and of course, note that these
correction factors are in addition to any bolometric correction that
should already have been applied to $E_{\rm rad}$, if $E_{\rm rad}$ is
derived from optical photometry, and they do not account for interior 
SN ejecta that have yet to reach the reverse shock).  This geometric efficiency factor is
significant, since SN~2015da is already a very energetic event before
correction.  A similar correction to the inferred energy was discussed
previously regarding the asymmetry in SN~2009ip
\citep{mauerhan14,smith14}, but in that case the total required
explosion energy was corrected from around $10^{50}$\,erg up to
$10^{51}$\,erg.  In the case of an SLSN~IIn like SN~2015da, this
correction pushes the limits more because, as we discuss below, the
observed value of $E_{\rm rad}$ already exceeds the canonical SN
explosion energy of $10^{51}$\,erg.

%% 4.2
\subsection{CSM interaction: mass and energy budget}

Earlier in Section 3.2, from the $R/r$-band light curve of SN~2015da,
we measured a value for the total integrated radiated energy of
$E_{\rm rad} = 1.6 \times 10^{51}$\,erg.  We estimated a somewhat
higher value of $E_{\rm rad} = 1.9 \times 10^{51}$\,erg if we adopted
the ``pseudo-bolometric'' correction from T20.  When we consider the
likely correction factors $f$ and $q$ for the conversion efficiency
and asymmetric geometry, respectively, the value of $E_{\rm exp} >
E_{\rm rad}/fq$ quickly climbs to around
(5--10) $\times 10^{51}$\,erg.  Thus, just from the radiated energy
budget generated by CSM interaction, we already have important
constraints on the explosion, because this is too much for any SN~Ia,
and it is on the high end for normal core-collapse SNe.  However, even this
is still a lower limit to the required explosion energy, because we
must also account for the kinetic energy in the SN ejecta that have
not yet reached the reverse shock, and the kinetic energy due to the
final coasting speed and swept-up mass in the post-shock shell.  The
former is difficult to estimate, requiring models to infer
the mass and speed of ejecta that have not yet reached the reverse
shock by $\sim$2600\,days.  However, observations do allow good
constraints on the mass and kinetic energy of the post-shock shell.

When CSM interaction dominates the observed luminosity, as is thought
to be the case in SLSNe~IIn, the luminosity can be used to infer the
density of the CSM and the progenitor's mass-loss rate using the
relation \cite[][]{smith17}

\begin{equation}
L = \frac{1}{2} w V_{\rm SN}^3 = \frac{\dot{M} V_{\rm SN}^3}{2 V_{\rm CSM}}\, ,
\end{equation}

\noindent where $w$ is the wind density parameter $w = 4 \pi R^2
\rho$ or $w = \dot{M} / V_{\rm CSM}$, $\dot{M}$ is the progenitor
star's mass-loss rate, and $V_{\rm SN} = V_{\rm CDS}$ is the speed of
the forward shock, indicated by the observed speed of the CDS.  Here, $L$ is the luminosity generated by CSM interaction;
this luminosity might be greater than the observed optical luminosity
$L_{\rm opt}$ (or even the pseudo-bolometric luminosity), if some
significant portion of the radiated energy escapes as X-rays, for
example.  In the case of SLSNe~IIn, we generally assume that $L
\approx L_{\rm opt}$ at early times during the main light-curve peak
when X-rays are absorbed and thermalised \citep{sm07}, but this
becomes an increasingly risky assumption at later times, when the
optical depth drops and significant X-ray flux can escape. Thus, once
again, one should be mindful of another correction factor that acts to
raise the mass and energy budget even more.

We adopt a constant value of 90\,km\,s$^{-1}$ for $V_{\rm CSM}$, based
on the steady narrow P~Cygni absorption seen throughout the evolution
of SN~2015da (Fig.~\ref{fig:narrow}).  The value for the speed at which the
shock sweeps into the CSM, $V_{\rm SN} = V_{\rm CDS}$, is provided
observationally by the line width of emission from the post-shock gas
in the CDS. This is indicated by the intermediate-width components of
emission lines, but only after their transition from Lorentzian
profiles (dominated by electron scattering in the CSM) to the profiles
after peak that trace expansion of post-shock gas.  As noted earlier, the red wings
of these emission lines suffer selective velocity-dependent
extinction, probably from internal dust formation (see below), so the
value of $V_{\rm CDS}$ is best indicated by 2 times the blue HWHM velocity, or the
FWHM of symmetric Gaussians that are only fit to the blue wing.  The value of
$V_{\rm CDS}$ is time dependent, because the shock decelerates over
time as it gets mass loaded. As noted earlier, observed estimates of
this velocity are plotted in Fig.~\ref{fig:vel}.  The value of
$V_{\rm SN} = V_{\rm CDS}$ that we adopt at each time step is given by
the smooth blue curve in Fig.~\ref{fig:vel}.

\begin{figure}\begin{center}
\includegraphics[width=3.0in]{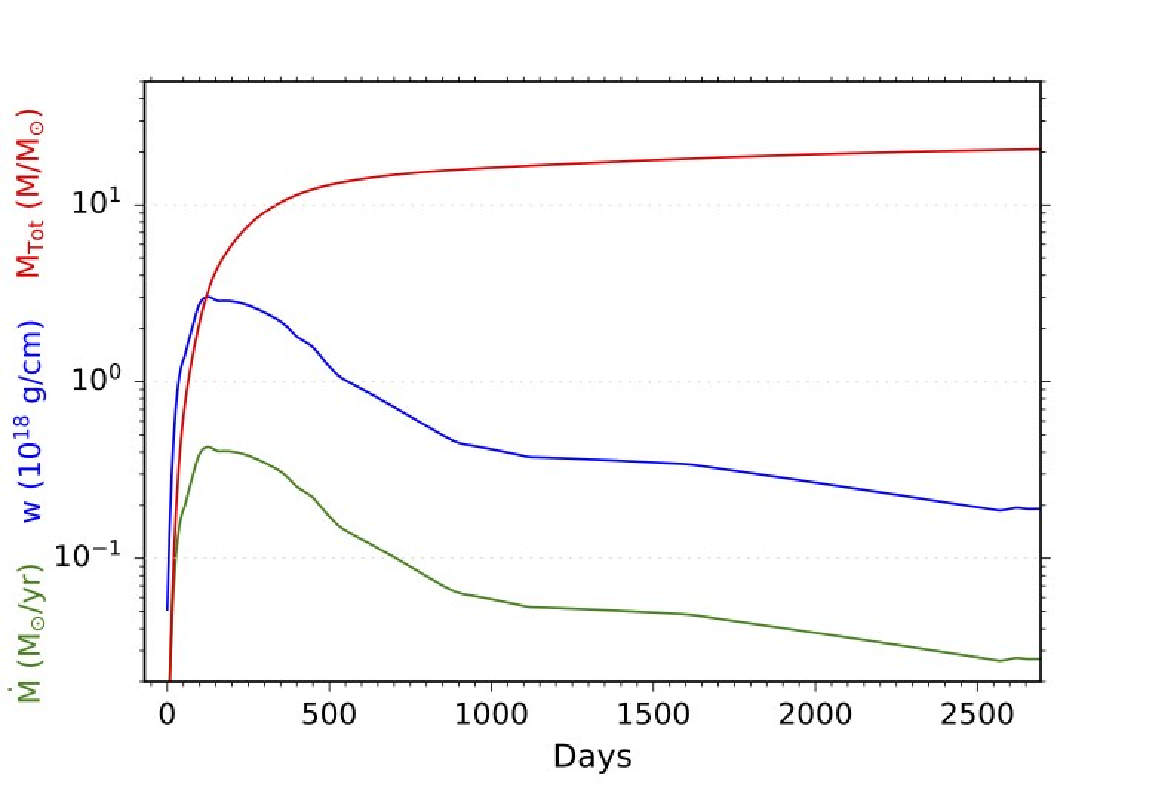}
\end{center}
\caption{Values for the progenitor mass-loss rate (green), the wind
  density parameter $w$ (blue), and the cumulative mass swept up by the
  shock $M_{\rm Tot}$ (red) as a function of observed time since explosion.  These are derived from the
  observed luminosity and the CSM and shock speeds (see text).}
\label{fig:mdot}
\end{figure}

With observed estimates for $L$, $V_{\rm SN}$, and $V_{\rm CSM}$, we
can calculate the characteristic wind density parameter and progenitor
mass-loss rate corresponding to the CSM overtaken by the shock, given
by

\begin{equation}
\dot{M} = w V_{\rm CSM} = 2 \ L \ \frac{V_{\rm CSM}}{(V_{\rm SN})^3}\, ,
\end{equation}

\noindent at each time step.  Calculated values for $w$ and $\dot{M}$ at each
time are plotted in blue and green (respectively) in
Figure~\ref{fig:mdot}.  The time in days on the horizontal axis is the
time since the SN exploded, $t_{\rm SN}$.  As noted in Section 3.7,
this can be converted to the timescale it took the progenitor to make
the CSM via $t_{\rm CSM} = t_{\rm SN} \times V_{\rm SN} / V_{\rm
  CSM}$.  The red curve, $M_{\rm Tot}$, shows the cumulative amount of
CSM mass that has been swept up by the shock at each time step; it
is the integral of $\dot{M} \times \Delta t_{\rm CSM}$.

From Figure~\ref{fig:mdot} we see that SN~2015da requires at least
20\,$M_{\odot}$ of swept-up CSM.  It therefore joins some of the most
extreme SLSNe~IIn like SN~2006gy, SN~2006tf, SN~2003ma, and SN~2016aps, which require
similar amounts of CSM
\citep{sm07,smith08tf,smith10,woosley07,rest11,nicholl20}.
Normally, one must be cautious with interpreting a value for $\dot{M}$
because the pre-SN mass loss can be more explosive than steady, but in
this case of SN~2015da, we observe a steady CSM expansion speed of 90\,km\,s$^{-1}$ throughout its evolution.  This indicates that the progenitor
was losing material at a relatively steady (but astonishing) rate of
$\sim 0.4$\,M$_{\odot}$\,yr$^{-1}$ for several decades before explosion,
and at a somewhat less extreme rate of $\sim 0.04$\,M$_{\odot}$\,yr$^{-1}$ for another 2 centuries before that.

The final coasting speed of the swept-up CDS also places important constraints on the mass of SN ejecta, since the SN ejecta must provide not only the energy needed for the SN luminosity, but also the momentum imparted to the accelerated CSM.  From simple momentum conservation, we have the constraint

\begin{equation}
M_{\rm e} = M_{\rm CSM}  \frac{(V_{\rm CDS} - V_{\rm CSM})}{(V_{\rm e} - V_{\rm CDS})}\, ,
\end{equation}

\noindent where, as before, $V_{\rm CDS}$ is the observed speed of the post-shock CDS, and $M_{\rm CSM}$ and $V_{\rm CSM}$ are the mass and speed of the pre-SN CSM.  Also, $V_{\rm e}$ is the effective velocity of the SN ejecta, and $M_{\rm e}$ is the mass of SN ejecta that has hit the reverse shock.  Since ejecta with $V_{\rm e} < V_{\rm CDS}$ have not yet contributed their momentum to the CDS, $M_{\rm e}$ is a lower limit to the required total SN ejecta mass.  Observations noted above provide values of $M_{\rm CSM} = 20$\,M$_{\odot}$, $V_{\rm CDS} \ge 2100$\,km\,s$^{-1}$, and $V_{\rm CSM} = 90$\,km\,s$^{-1}$.  We do not know $V_{\rm e}$ {\it a priori}, but for a typical SN~Ia, the velocity $\sqrt{2E/M}$ is about 8500\,km\,s$^{-1}$.  These values would then require SN ejecta mass of $M_{\rm e} > 6.5$\,M$_{\odot}$.  This clearly rules out any SN~Ia scenario for SN~2015da, which is in good agreement with the total CSM mass budget exceeding 20\,M$_{\odot}$ and the total energy well above $10^{51}$\,erg, both of which also rule out
any model involving SNe~Ia.  

What mechanism powers the extreme pre-SN mass loss?  As noted earlier, the timescale of a few
centuries before explosion rules out wave driving as the source,
% of the extreme pre-SN mass loss, 
since this only works on
timescales around 1\,yr \citep{qs12,sq14,fuller17}.  Recent work also suggest that this mechanism may inject less power into the envelope than previously thought, and may favour lower-mass progenitors \citep{wf21}, exacerbating the difficulty for this mechanism to explain SLSN~IIn precursor mass loss.

The only remaining proposed mechanisms for extreme pre-SN mass loss
include the pulsational pair instability or other nuclear burning
instabilities \citep{woosley07,woosley17,sa14,am11,renzo20}, or pre-SN
binary interaction \citep{sa14,schroder20,fw98}.  

Of these,
pulsational pair instability is the mechanism with the most clear and
well-established predictions.  This mechanism seems well suited to
account for the inferred CSM mass, the energy budget
of the pre-SN mass loss, and potentially the timescale for the
pre-SN mass-loss.  While the pulsational pair instability generally
occurs during O burning --- which, like wave driving, usually only
occurs around 1\,yr before death --- the energetic pulses instigated by
explosive O burning can expand the core, which relaxes on a long
thermal timescale.  As such, extreme mass loss can potentially occur
for many decades or even centuries before the star finally dies \citep{woosley17}.
However, the fact that the strong mass loss from pulsational pair
instability occurs via a series of hydrodynamic pulses means that it
is driven off the star by shocks, and thus results in fast
expansion speeds.  The bulk expansion speeds of the ejecta are always
around 2000\,km\,s$^{-1}$ or larger \citep{woosley17,renzo20,ws22}.
Moreover, the first pulse generally removes the residual H envelope,
with the subsequent ejecta being H poor \citep{renzo20,ws22}.  These
are robust predictions, and they strongly disfavour pulsational
pair instability as the trigger for the strong pre-SN mass loss in
SN~2015da.  This is because SN~2015da showed a remarkably steady CSM
expansion speed of only 90\,km\,s$^{-1}$ for centuries before explosion,
and always exhibited strong H lines, even in the broad lines from the
freely expanding SN ejecta.  This is difficult to achieve with a
series of strong pulses that suddenly eject the star's H envelope.  The
observational expectations for other nuclear burning instabilities are
not as clear as for the pulsational pair instability, so further work
on these is warranted.  Since the energy is deposited deep inside the
star and should steepen to a shock as it approaches the surface
\citep{fr18}, the observational consequences might be similar to those
of the pulsational pair instability.

On the other hand, a slow and steady expansion speed of around 90\,km\,s$^{-1}$ over centuries may be well suited to the slow leaking of mass
that must occur during the relatively long inspiral phase before a
merger event.  This roughly matches the $\sim$100\,km\,s$^{-1}$
pre-eruption mass loss seen in light-echo spectra of $\eta$~Carinae
before its merger \citep{smith18}, which also produced a CSM mass of order 20\,M$_{\odot}$ \citep{smith03mass}. Moreover, the fact that the
pre-explosion mass loss inferred for SN~2015da ramps up with time (see
Fig.~\ref{fig:mdot}) also agrees well with a merger scenario.  The
total expansion kinetic energy of the CSM ($0.5 \times 20$\,M$_{\odot} \times$ (90\,km\,s$^{-1}$)$^2$) is around $2 \times 10^{48}$\,erg, which
is easily supplied by the available orbital energy in a massive-star
merger.  So far, some sort of pre-SN merger event \citep{sa14} seems
like the most favourable explanation for providing the astounding
pre-SN mass loss for SN~2015da.  We note that while the pre-SN mass
loss of SLSNe~IIn closely resembles the strong mass loss of LBVs
\citep{smith07}, it is quite possible that most giant eruptions of
LBVs are themselves stellar merger events, as is thought to be the
case for $\eta$~Carinae \citep{smith18}.  The reason a
merger might occur immediately before a very energetic SN explosion is
still unclear, however.  Perhaps late burning phases cause the star to
swell, synchronising the binary interaction with core collapse
\citep{sa14}, or perhaps a merger with a compact companion triggers a
violent energetic explosion powered by accretion
\citep{fw98,schroder20,sa14}.  Further theoretical exploration in this
direction (including the predicted observational signatures) is
certainly worthwhile.

%% 4.3
\subsection{Blueshifted line profiles and dust}

SN~2015da also joins a number of well-observed SLSNe~IIn and SNe~IIn
that exhibit asymmetric blueshifted profiles in their
intermediate-width emission lines from the post-shock CDS.
Other well-studied examples include SN~2017hcc \citep{smith20},
ASASSN~15ua \citep{dickinson23}, KISS15s \citep{kokubo19}, SN~2013L \citep{andrews17}, SN~2010jl
\citep{smith12,gall14}, SN~2010bt \citep{eliasrosa18}, SN~2007rt
\citep{trundle09}, SN~2007od \citep{andrews10}, SN~2006tf
\citep{smith08tf}, SN~2005ip
\citep{smith09sn05ip,fox10,smith17sn05ip}, and SN~1998S
\citep{ms12,pozzo04}. This blueshift is also seen clearly in some
interacting H-poor SNe, most notably in SN~2006jc \citep{smith08jc}.

%-------------------

In most interacting SNe that show these blueshifted line profiles, the
lines begin symmetric (usually with Lorentzian-shaped profiles) and
evolve to become blueshifted over time. This, and the narrow
lines centred on zero velocity, argue against a pervasive tendency
for SN~IIn progenitors to launch one-sided CSM preferentially toward
Earth.  Instead, it is more likely that something within the SN
explosion is blocking the redshifted portions of the emitting
material.

The most common explanation for the observed systematic blueshift in
emission-line profiles in interacting SNe is that new dust
grains are forming, either in the SN ejecta or in the dense rapidly
cooling post-shock CDS.  Only dust that is found internal
to the SN (i.e., not circumstellar) can preferentially block the
receding material, simply because the emission from the far side has a
long path length passing through the explosion interior,
whereas the blueshifted emission does not.  This effect was most
famously seen in SN~1987A, where developing blueshifted line
profiles from SN ejecta were accompanied by an increased rate of fading
and by growing IR excess emission
\citep{danziger89,lucy89,gn90,wooden93,colgan94}.  Models of the line-profile evolution indicate continual growth of dust grains for decades
in  SN~1987A \citep{bevan16}.  

Each of these observational effects --- increased rate of fading, excess IR
 emission, and blueshifted line profiles --- taken on their own
might be ambiguous, because there may be more than one cause for each.
But when all three developments occur together, as in SN~1987A, it
presents a strong case that new dust grains are forming in the SN.
The first interacting SN to clearly show all three signs was SN~2006jc
\citep{smith08jc}, which also exhibited an outburst in X-ray emission
\citep{immler08} accompanied by He~{\sc ii} $\lambda$4686 emission at the same
time.  This coincidence, the early onset of these effects at $\sim$50\,days, and the fact that the intermediate-width He~{\sc i} lines showed
a deficit of flux at zero velocity as well as on the red wing,
indicated clearly that the location of the newly formed dust was in the
post-shock region within the CDS, and not in the fast SN ejecta.
\citet{smith08jc} noted that this is quite similar to the 
episodic post-shock dust formation that occurs in eccentric 
colliding-wind binaries like WR~140 \citep{hgg79,williams90,monnier02} and
$\eta$~Car \citep{smith10eta}, where episodes of post-shock dust
formation at periastron are accompanied by strengthening X-ray
emission and He~{\sc ii} $\lambda$4686 emission from the shock.

In SNe~IIn, the late-time luminosity is typically dominated by ongoing
CSM interaction rather than radioactive decay, so unfortunately, it is
difficult to tell if there is an increased rate of fading from
additional extinction (because we don't really know the intrinsic luminosity
decline rate, as we do for $^{56}$Co decay).  As such, we are left with
the two remaining indicators of dust formation: IR excess emission
and blueshifted line profiles.

Mid-IR excess emission from dust is seen in every example mentioned
above where SNe~IIn or Ibn show blueshifted line profiles (when 
mid-IR observations were obtained).  From an
IR excess alone, however, it is unclear where the dust is located or
if it is new.  It could be formed in the SN ejecta or in the
post-shock CDS, but in principle the dust might also be pre-existing and could be
seen as an IR echo \citep{gerardy02}.  This echo could be a true echo
where CSM dust is heated by the main peak of light curve, or
a continual echo where luminosity from ongoing CSM interaction
heats CSM dust just ahead of shock.  In fact, regardless of whether
new dust forms in the SN, it is likely for any SN~IIn to have an IR
echo, since the very dense and slow CSM that is required to make it
appear as an SN~IIn is also likely to be dusty.  The IR excess emission
in both SN~2010jl \citep{andrews11,sarangi18}\footnote{For SN~2010jl, \citet{sarangi18} concluded that new post-shock dust formation dominates the IR emission after day 380.}  and SN~2015da (T20) were
interpreted as IR echoes.  Most SNe~IIn show continually re-heated IR
echoes from ongoing late-time CSM interaction \citep{fox13}.  The
presence of an IR echo does not, of course, preclude the possibility
that new dust also forms in the SN.

As noted above, several SNe~IIn have shown clear evidence for a net
blueshift in the centroid of emission lines, and in many cases, an
asymmetric skewed line-profile shape.  As in the case of SN~2006jc,
the specific velocity components of the lines that show the blueshift,
plus the detailed shape of the line, can help us decipher the region
where new dust must be forming.  In SN~1987A, the blueshift was seen
in the broad emission lines from the freely expanding SN ejecta.  In
SNe~IIn and Ibn, the blueshift is sometimes seen in the broad ejecta
components, but is more commonly seen in the intermediate-width components
emitted by the post-shock gas
\citep{smith09sn05ip,smith12,gall14,smith20}.  Modeling of the line
profiles in SNe~IIn has confirmed that dust formation in the ejecta or
post-shock regions (or sometimes both) can account for the observed
line profiles, and models can disentangle the relative amount in each
region \citep{chugai18,bevan19,bevan20}.  Dust that forms only in the SN ejecta
can block the red wings of both the broad components and potentially
the intermediate-width component, since the most redshifted portion of
the CDS  may be behind the inner SN ejecta.
However, dust in the inner SN ejecta cannot block emission from
the CDS at around zero velocity, since this material is moving in the
plane of the sky and is outside the SN ejecta; when intermediate-width
components show a deficit of emission in both the red wings and around
zero velocity, this requires dust in the CDS
\citep{smith08jc,smith12,chugai18}.

SN~2015da presents another case of an SLSN~IIn that clearly showed both
the IR excess from warm dust and blueshifted emission-line profiles.
T20 interpreted the IR excess and the blueshifted profiles as separate
phenomena: the IR excess was attributed to an IR echo from
pre-existing CSM dust, as noted above, whereas the pronounced
blueshift was attributed to the same scenario proposed for SN~2010jl
by \citet{fransson14}.  \citet{fransson14} attributed the blueshifted
profiles in SN~2010jl at all epochs to electron scattering of emission
from pre-shock asymmetric CSM gas that was mostly on our side of the
SN, and was accelerated toward the observer by radiation from the SN
shock, thus producing intermediate-width lines with a symmetric shape,
but with a centroid offset from zero velocity.  As discussed in detail
by \citet{smith20}, however, there are three key reasons why the model
proposed by \citet{fransson14} and adopted by T20 does not work.  (1)
the narrow emission components from the unshocked CSM are in fact
detected, but they are not blueshifted, and they are not at the centre
of the blueshifted intermediate-width component.  Moreover, the narrow
components show narrow P~Cygni profiles that indicate the same slow
velocities as the narrow emission.  This means that the CSM has not
been accelerated toward us, and that the blueshifted intermediate-width components
cannot result from broadening of the observed narrow emission.  (2) Although
acceleration of pre-shock CSM may occur, the observed amount of a few
hundred km\,s$^{-1}$ is too large \citep{dessart15}.  Moreover, any
radiative acceleration of CSM should be strongest when the SN
luminosity is the highest (i.e., at peak), and should diminish at late
times when the luminosity falls.  But observations show the opposite,
where the blueshift persists until late times and becomes even more
pronounced with time after peak.  (3) Similarly, the hypothesis where electron
scattering dominates the line broadening requires high electron
scattering optical depths, and this is expected to diminish with time
as the SN fades and the electron-scattering opacity drops.  But
instead, the intermediate-width components tend to become even more blueshifted
as the continuum opacity disappears, ruling out the electron-scattering
model.  Surely electron scattering does broaden the wings at early times when the
lines are symmetric Lorentzians, but it cannot account for the
asymmetric blueshifted profiles at late times.

A possible alternative cause of blueshifted profiles is that the continuum
photosphere in the SN itself, rather than dust, blocks some of the
redshifted emission. This was discussed in the context of
the blueshifted profiles in SN~2010jl by \citet{smith12}, who noted
that as the SN fades and the continuum photosphere transitions to
optically thin CSM interaction, this effect ceases to work and the
line profiles should become symmetric again.  In SN~2015da and previous
examples like SN~2010jl and SN~2017hcc, observations show the blueshift
persisting until very late phases more than 1000\,days after explosion.
Thus, while this occultation by the SN photosphere may help explain
some of the early blueshift, it cannot explain the evolution after peak, so
dust formation is required.

%-------------------

Besides the IR excess and the blueshifted profiles, there is one
additional tell-tale sign of dust formation within the SN.  When the
blueshifted line profiles are seen, they may also exhibit a wavelength
dependence, with stronger blueshifted asymmetry at shorter
wavelengths.  For example, compared to the asymmetric blueshifted
profile seen in H$\alpha$, H$\beta$ might show even more
blueshift, whereas a near-IR line like Pa$\beta$ might have a more
symmetric profile.  If this effect is seen, it leaves little doubt that
new dust formation is the culprit behind the blueshifted profiles,
because dust causes greater extinction at shorter wavelengths, whereas
electron-scattering opacity is wavelength independent.  Such
wavelength dependence to the blueshift is indeed observed in some SLSNe~IIn
when good multiwavelength data are available, as seen clearly in both
SN~2010jl \citep{smith12,maeda13,gall14} and SN~2017hcc \citep{smith20}.  These cases indicate the rapid formation of relatively large grains \citep{gall14,smith20}.  This
evidence requires
either IR spectra with good S/N, or blue spectra covering
H$\beta$ with very high S/N (H$\beta$ is much fainter than
H$\alpha$ in interacting SNe at late times).

Considering all these points above, we find that in SN~2015da ---
as in the previous cases of SN~0210jl and SN~2017hcc --- it is clear
that the observed asymmetric blueshifted profiles arise because of
copious new dust formation in the post-shock CDS.  For the first few
epochs leading up to the time of peak luminosity, the
intermediate-width profiles of H$\alpha$ are indeed Lorentzian in
shape and symmetric, centred on zero velocity.  After peak, however,
when the blueshift becomes strong, the line wings are no longer
Lorentzian in shape.  They become complex asymmetric blueshifted
profiles, with centroids that are not just offset from zero velocity,
but also profiles skewed in shape and multicomponent.  The
centroids of the intermediate-width components are shifted to the blue
and offset from the narrow lines, which remain close to zero.  This offset of the centroid of the
intermediate-width component without any similar offset in the
velocity of the narrow component definitively rules out electron
scattering of radiatively accelerated CSM as the explanation for
the profiles.  Additionally, the blueshift becomes more pronounced as
the SN fades, and remains strong until late times.  This is the
opposite of expectations for electron scattering and radiative
acceleration, but it is expected for dust, which continues to form as
the material expands and cools \citep{gall14,li22}, and where dust opacity remains high
even at low temperatures.

While we were not able to obtain high-quality IR spectra of SN~2015da, we do see
evidence for some wavelength dependence to the blueshift.  We only have
a few epochs of spectra with broad wavelength coverage that includes
H$\beta$ (see Fig.~\ref{fig:spectra}), and in many of these, the S/N around H$\beta$ and the presence of other emission lines
complicates a comparison.  But in our last epoch obtained with the LBT,
H$\beta$ is detected with good S/N; this profile is shown
as a blue dashed curve at the bottom of Figure~\ref{fig:ha.dust}.
When we adjust the flux so that the blue wings of H$\beta$ and H$\alpha$ match,
there is clearly an additional deficit of flux on the red side of
H$\beta$ as compared with H$\alpha$, even though H$\alpha$ is itself
already blueshifted and asymmetric.  The difference between H$\beta$
and H$\alpha$ at this epoch is similar to the differences between these
two lines at late times in SN~2010jl and SN~2017hcc
\citep{smith12,gall14,smith20}. While this is admittedly only one epoch,
it is at least consistent with the hypothesis that dust is causing
wavelength-dependent extinction of the redshifted emission at this late
epoch.

Additionally, we noted above that the blueshift became strongest in
SN~2015da around 400--800\,days after explosion.  This is roughly the
same time when the IR excess in SN~2015da was observed to be the
strongest (T20). We cannot rule out the possibility that some
portion of the IR excess is due to an IR echo from pre-existing dust,
but this coincidence argues that a good portion of it may also be due to new
dust forming in the SN.

Where is the new dust located in SN~2015da?  The blueshifted
intermediate-width profiles require that some new dust is forming in
the post-shock CDS or in the SN ejecta (or both).  Recall that, in
principle, dust within the SN ejecta may block some of the redshifted
emission from the back side of the CDS, and thus suppress the red wings
of even the intermediate-width components.  There are two indications,
however, that the new dust formation in SN~2015da is primarily in the
post-shock CDS.  First, the flux deficit on the red side of the lines
begins right at zero velocity, and there may even be some loss of flux
around zero velocity.  Second, and more clearly, we do not see any
deficit of flux on the red wing of the broad (7000\,km\,s$^{-1}$)
emission components.  The broad H$\alpha$ line wings are symmetric
(Fig.~\ref{fig:ha.fits}).  Since these broad emission wings are
tracing high-speed material in the freely expanding SN ejecta, and
since they show no asymmetry, it is unlikely that there is a
significant amount of obscuring dust in the SN ejecta that blocks the
red side.  Therefore, obscuring dust that affects the
intermediate-width components must be primarily in the post-shock zones
of the CDS.  This new dust may be new dust that has
condensed purely from the gas phase, or it may arise from CSM dust that
was incompletely destroyed by the forward shock and has regrown in the
rapidly cooling post-shock shell.

%% Conclusion
\section{Conclusion} \label{conclusion}

We present photometry and spectroscopy of SN~2015da, an unusual SLSN~IIn that was discovered very soon after explosion, had a long rise to peak brightness, and was extraordinarily luminous and long-lasting.  Our data cover times up to $\sim$3000\,days post explosion.  Overall, it is similar to the previous energetic events SN~2003ma \citep{rest11} and SN~2016aps \citep{nicholl20}.  SN~2015da was already discussed in the literature by T20, and our data are complementary to the data presented in that study, although extended to later times.  Our interpretation differs from T20 in a few respects, especially concerning the evolution of emission-line profiles. Key results from our study of SN~2015da are as follows.

(1) Integrating the observed $R/r$-band light curve up to about day 2600 yields a total radiated energy of at least $E_{\rm rad} = 1.6 \times 10^{51}$\,erg (1.6\,FOE), and with a modest bolometric correction this rises to at least 1.9\,FOE (not including X-rays). 

(2) The integrated value of $E_{\rm rad}$ is, of course, only a lower limit to the explosion kinetic energy.  Accounting for the typical efficiency of CSM interaction, and including a correction for the asymmetric geometry of the CSM indicated by polarization \citep{bilinski23} --- plus the fact that some portion of the fast SN ejecta have not yet hit the reverse shock, and the escape of X-ray luminosity at late times ---  the explosion kinetic energy was likely to be 5--10 FOE or more.  This rules out a Type Ia SN model for SN~2015da, and also seems more energetic than typical core-collapse SNe.  The momentum of the swept up CDS also requires a SN ejecta mass $>$6.5 $M_{\odot}$, also clearly ruling out any SN Ia scenario.

(3) Even at its faintest point $\sim$8\,yr after explosion, SN~2015da remains as luminous as the peak of a typical SN~II-P.  In order to power the long-lasting high luminosity with CSM interaction requires a total CSM mass of more than 20\,M$_{\odot}$.  This mass was lost by the progenitor in the last $\sim$200\,yr before explosion.  

(4) If this CSM mass corresponds to an ejected H envelope, then accounting for the likely mass of the SN ejecta plus mass lost by the progenitor throughout its evolution via stellar winds, the total mass budget requires a very massive progenitor star with $M_{\rm ZAMS}$ of at least 50\,M$_{\odot}$. (This is the equivalent initial mass of a single star; obviously the initial mass could have been somewhat less than this if the progenitor evolved as a binary and gained mass or merged during in its evolution.)

(5) The mass-loss rate that produced the CSM needed to power the light curve ramped up from $\sim$0.04\,M$_{\odot}$\,yr$^{-1}$ at 200\,yr before core collapse, up to $\sim$0.4\,M$_{\odot}$\,yr$^{-1}$ in the decades before explosion.  This $\dot{M}$ is much higher than any normal stellar wind can produce.  

(6) Moderate-resolution spectra reveal a persistent narrow emission component arising from pre-shock gas in the CSM.  It changes very little over 3000\,days and displays a P~Cygni profile at most epochs.  This indicates a relatively constant and slow CSM outflow speed of $\sim$90\,km\,s$^{-1}$. Such a speed is very fast for RSG winds, and on the slow end for an LBV, but it is similar to outflow speeds observed in the equatorial L2 mass loss from contact binary systems.

(7) Evolution of intermediate-width emission-line profiles clearly shows the blueshifted asymmetry that is often seen in SNe~IIn.  Emission lines exhibit a symmetric Lorentzian profile at early times up to peak luminosity, but become asymmetric and blueshifted around peak and afterward, and this blueshift persists until the last observed epoch.  For a number of reasons, this blueshift must be caused by the formation of new dust grains within the SN.  A similar blueshifted asymmetry is {\it not} seen in the broad emission component arising from the freely expanding SN ejecta, indicating that the location of the new dust formation must be in the post-shock shell (the CDS).  Dust grains appear to be forming continually from $\sim$110\,days to our last observation around day 3030.

\section*{Acknowledgements}
%Special thanks to  
Support was provided by NSF grants AST-131221 and AST-151559. 
A.V.F.'s supernova group at U.C. Berkeley has received financial assistance from the Christopher R. Redlich Fund, Kathleen and Briggs Wood, and many other individual donors.
R.\ Hofmann assisted with early phases of the data collection and analysis, and we thank J.\ Mauerhan and M.\ Graham for help with the Lick/Kast observations.
%A major upgrade of the Kast spectrograph on the Shane 3\,m telescope at Lick Observatory was made possible through generous gifts from William and Marina Kast as well as the Heising-Simons Foundation. 
Research at Lick Observatory is partially supported by a generous gift from Google.
 Observations using Steward
Observatory facilities were obtained as part of the large observing
program AZTEC: Arizona Transient Exploration and Characterization.
Some of the observations reported in this paper were
obtained at the MMT Observatory, a joint facility of the University
of Arizona and the Smithsonian Institution.

This research was also
based in part on observations made with the LBT. The LBT is an
international collaboration among institutions in the United States,
Italy and Germany. The LBT Corporation partners are the University
of Arizona on behalf of the Arizona university system; the Istituto
Nazionale di Astrofisica, Italy; the LBT Beteiligungsgesellschaft,
Germany, representing the Max-Planck Society, the Astrophysical
Institute Potsdam, and Heidelberg University; the Ohio State University
and the Research Corporation, on behalf of the University of
Notre Dame, University of Minnesota, and University of Virginia.

Some of the data presented herein were obtained at the W. M. Keck
Observatory, which is operated as a scientific partnership among
the California Institute of Technology, the University of California
and NASA; the observatory was made possible by the generous
financial support of the W. M. Keck Foundation. The authors wish
to recognise and acknowledge the very significant cultural role and
reverence that the summit of Maunakea has always had within the
indigenous Hawaiian community. We are most fortunate to have the
opportunity to conduct observations from this mountain.
We thank the staffs at the MMT, Lick, and Keck Observatories for
their assistance with the observations.

\section*{Data Availability}

The data underlying this article will be shared on reasonable request
to the corresponding author.

\bibliographystyle{mnras}
\bibliography{PSNJ13522}

%\begin{thebibliography}{99}
%\end{thebibliography}
%%%%%%%%%%%%%%%%%%%%%%%%%%%%%%%%%%%%%%%%%%%%%%%%%%

%%%%%%%%%%%%%%%%% APPENDICES %%%%%%%%%%%%%%%%%%%%%
%\appendix

%%%%%%%%%%%%%%%%%%%%%%%%%%%%%%%%%%%%%%%%%%%%%%%%%%
% Don't change these lines
\bsp	% typesetting comment
\label{lastpage}
\end{document}